\documentclass[aps,prm,twocolumn,floatfix,amsmath,nofootinbib]{revtex4-1}

\usepackage{graphicx}
\usepackage[dvipsnames]{xcolor}
\usepackage{dsfont}
\usepackage[normalem]{ulem}
\usepackage{dcolumn}
\usepackage{multirow}
\usepackage{textcomp}
\usepackage{placeins}
\usepackage{subfigure}
\usepackage{nicefrac}
\usepackage{xfrac}
\usepackage{float}
\usepackage{warpcol}
\usepackage[utf8]{inputenc}
\hfuzz=\maxdimen
\newcommand{\eg}{e.g.,\@ }
\newcommand{\ie}{i.e.,\@ }

\newcommand{\etal}{\textit{et al.\@ }}
\newcommand{\abinitio}{\textit{ab initio}}

\newcommand{\vcn}[1]{\hat{#1}}

\newcommand{\vc}[1]{\boldsymbol{\mathrm #1}}
\newcommand*\diff{\mathop{}\!\mathrm{d}}

\newcommand{\abs}[1]{\lvert#1\rvert}
\DeclareMathOperator{\W}{\mathrm{W_0}}

\begin{document}

\title{Material systems for FM-/AFM-coupled skyrmions in Co/Pt-based multilayers}
\author{Hongying Jia}
\email[Corresponding author:\@ ]{h.jia@fz-juelich.de} 
\author{Bernd Zimmermann}
\author{Markus Hoffmann}
\author{Moritz Sallermann}
\author{Gustav Bihlmayer}
\author{Stefan Bl\"ugel}
\affiliation{Peter Gr\"unberg Institut and Institute for Advanced Simulation, Forschungszentrum J\"ulich and JARA, 52425 J\"ulich, Germany}

\begin{abstract}

Antiferromagnetically coupled magnetic skyrmions are considered ideal candidates for high-density information carriers. This is due to the suppressed skyrmion Hall effect compared to conventional skyrmions and a smaller size due to the cancellation of some contributions to the  magnetostatic dipolar fields. By means of systematic first-principles calculations based on  density functional theory we search for suitable materials that can host antiferromagnetically coupled skyrmions. We concentrate on fcc-stacked (111)-oriented metallic $Z$/Co/Pt ($Z=4d$ series:  Y--Pd, the noble metals: Cu, Ag, Au, post noble metals: Zn and Cd)  magnetic multilayers of films of monatomic thickness. We present quantitative trends of magnetic properties: magnetic moments, interlayer exchange coupling,  spin stiffness, Dzyaloshinskii-Moriya interaction, magnetic anisotropy, and the critical temperature.  We show that some of the $Z$ elements (Zn, Y, Zr, Nb, Tc, Ru, Rh, and Cd) can induce antiferromagnetic interlayer coupling between the magnetic Co layers, and that they influence the easy magnetization axis. Employing a multiscale approach, we transfer the micromagnetic parameters determined from \abinitio\ to a micromagnetic energy functional and search for one-dimensional spin-spiral solutions and two-dimensional skyrmions. We determine the skyrmion radius by numerically solving the equation of the skyrmion profile. We found an analytical expression for the skyrmion radius that covers our numerical results and is valid for a large  regime of micromagnetic parameters. Based on this expression we have proposed a model that allows to extrapolate from the \abinitio\ results of monatomic films to multilayers with Co films consisting of several atomic layers containing 10-nm skyrmions. We found thickness regimes where tiny changes of the film thickness may alter the skyrmion radius by orders of magnitude. We estimated the skyrmion size as function of temperature and found that the size can easily double  going from cryogenic to room temperature.  We suggest promising material systems for ferromagnetically and antiferromagnetically coupled spin-spiral and skyrmion systems.

\end{abstract}

\maketitle
\section{Introduction}
The skyrmion is a mathematical concept first formulated in the early 1960s by the British physicist Tony Skyrme~\cite{Skyrme:61.1,Skyrme:61.2,Skyrme:62}. It describes topological solitons which are solutions of a nonlinear field equation. The solitonic nature is reflected in their existence as discrete entities emerging from a continuous background field. These entities,  localized in space,  behave like particles and can be conceived as stable geometric field twists, which  cannot be unwound by continuous transformations of the field. As conjectured by Bogdanov \etal\cite{Bogdanov:89},  the skyrmions materialized as chiral magnetic field configurations in magnets, first in  bulk crystals~\cite{Muhlbauer:09,Neubauer:09,Pappas:09} in 2009 and then 2011 in interface geometry~\cite{Heinze:11}. They are stabilized by the chiral symmetry-breaking Dzyaloshinskii-Moriya interaction (DMI) that emerges in magnetic materials with broken inversion symmetry and finite spin-orbit interaction.   

The notion of nanometer-size magnetic particles sparked  prospects in spintronic applications, for example, as information carriers of high storage density  or logical devices and recently also for neuromorphic and stochastic computing taking advantage of high mobility and low-energy consumption~\cite{Fert:13,Fert:17,Zhang:15.1,Bourianoff:18}. In order to turn an idea into a realization, materials are needed that allow 10-nm skyrmions at room temperature with long lifetimes, and enable low-energy and fast transport properties. Thinking of a technological realization, ferromagnetic Co/Pt-based magnetic multilayers (MMLs) are a promising materials platform in various respects~\cite{Park:NAM:2018,Yu:20,Sugimoto:19,Palermo:20,Gusev:20,Ma:20,Vetrova:20,Toneto:20,Sugimoto:20}. For example, the magnetization of Co is out of plane, the large spin-orbit interaction of Pt generates a sizable DMI, and individual sub-100-nm skyrmions at room temperature have been found experimentally in various Co/Pt systems~\cite{Moreau-Luchaire:16,Boulle:16,Woo:16,Pollard:17}. 

Despite these proof-of-principles experiments, the implementation of the concept of skyrmions as carriers of information in the context of a technology still has its limits. One of these is the size of the skyrmion, a second one the peculiar transport property of skyrmions, known as skyrmion Hall effect (SkHE):\@ the skyrmion trajectory deviates from the driving electric current direction and the skyrmion can even annihilate by touching the sample edges~\cite{Zang:11,Nagaosa:13,Everschor-Sitte:14,Jiang:17,Litzius:17}. 

\begin{figure}[t]
\centering
\includegraphics[width=80mm]{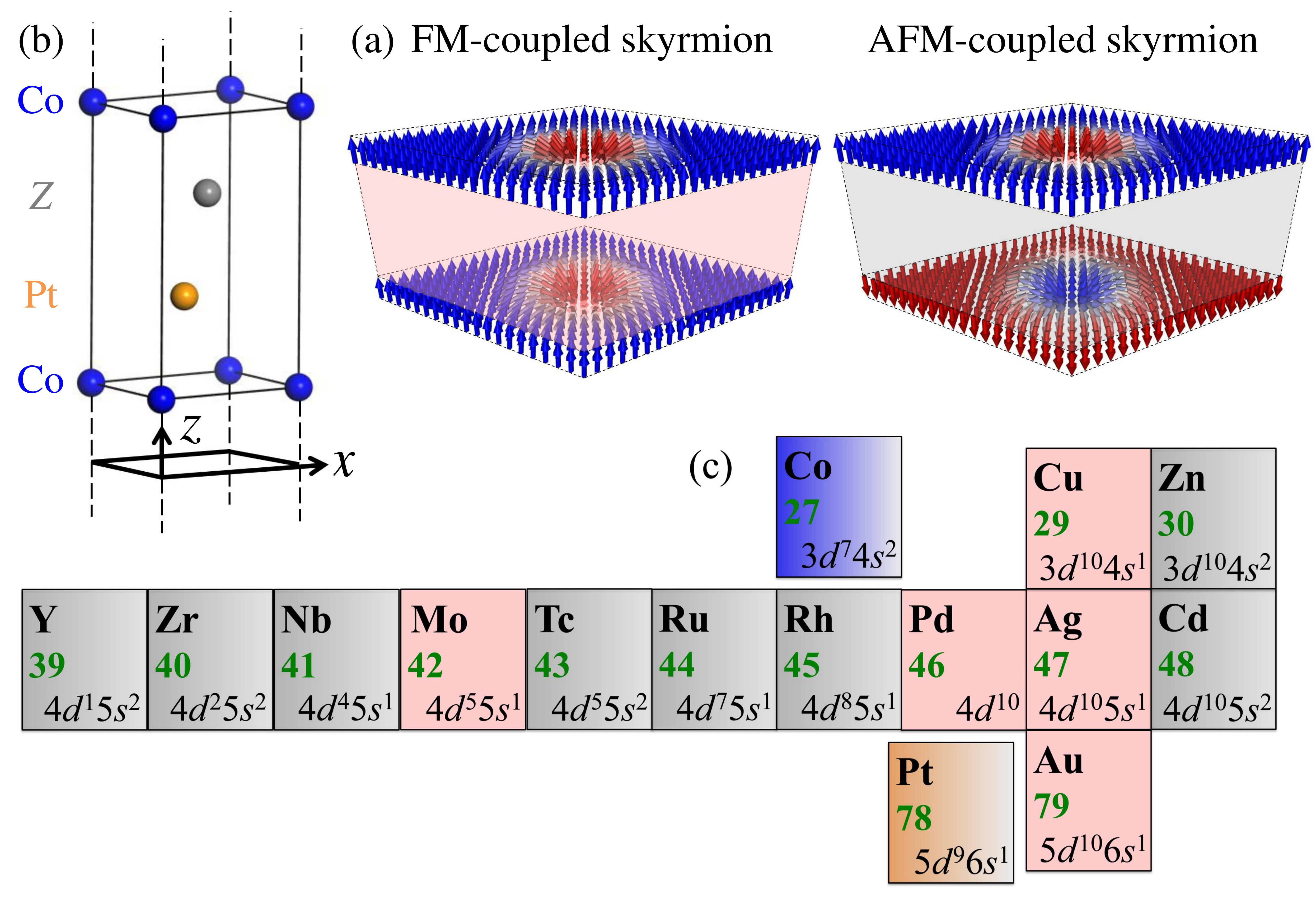}
\caption{(a) Schematic representation of ferromagnetic (FM) and antiferromagnetic (AFM) coupled skyrmions in the MMLs plotted using SPIRIT~\cite{spirit,muller:19}. (b) Sketch of crystalline structures of $Z$/Co/Pt ($Z=3d$:\@ Cu, Zn; 4$d$:\@ Y--Cd; 5$d$:\@ Au) MMLs. (c) Electronic configurations of the investigated transition-metal (TM) atoms. Numbers in green stand for their atomic number. The $Z$ atoms inducing a  SAF (FM) interlayer exchange coupling between magnetic Co layers are indicated by gray (peach) color.
}
\label{fig:structure}
\end{figure}

Synthetic antiferromagnetic (SAF) multilayers~\cite{Duine:18,Yang:19,Han:19,Li-Y:20}, \ie ferromagnetic (FM) films separated by an ultrathin antiferromagnetic coupling interlayer, offer a promising platform  to resolve these issues associated with skyrmions in ferromagnetic MMLs. Recently, antiferromagnetically (AFM) coupled magnetic bilayer skyrmions [see Fig.~\ref{fig:structure}(a)], a coupled pair of skyrmions with opposite topological charge ($Q=\mp 1$)  and polarity $p=\mp 1$ in synthetic antiferromagnets  have been proposed theoretically to reduce the dipolar stray field in order to make the skyrmions shrink in size and to  suppress the SkHE by cancellation of the Magnus force~\cite{Zhang:16.1,Zhang:16.2,Barker:16}. Subsequent theoretical papers on AFM-coupled skyrmions in SAF multilayers reported for example on: the method for efficient in-line injection of skyrmions~\cite{Gan:18}, the manipulation of AFM-coupled skyrmions~\cite{Koshibae:17,Loreto:19}, and the method to detect the emergence of AFM-coupled skyrmions~\cite{Buhl:17}. The interest in skyrmions in SAF-MMLs surged after the experimental observation of  individual 10-nm-size AFM-coupled skyrmions at room temperature by Legrand \etal \cite{Legrand:20}. Although compared to ferromagnetic skyrmions~\cite{Tomasello:17,Ang:19,Dohi:19,Zhou:20,Salimath:20} some advantages have theoretically been predicted for AFM-coupled skyrmions, such as higher speed, smaller size and more stability, and it has been shown experimentally that skyrmions in SAF have the potential to be small and stable, material systems that could host AFM-coupled skyrmions with optimal properties are still rare. At present, ruthenium is typically used to experimentally induce AFM coupling in Co-based multilayers~\cite{Legrand:20,Li-Y:20,Buttner:18,Kolesnikov:18,Karayev:19}. Therefore, the search for SAF materials that can accommodate AFM-coupled skyrmions is an important task in skyrmion research.

In this work, we report on a systematic state-of-the-art density functional theory (DFT) study of the magnetic properties of (111)-oriented, fcc (ABC) stacked  periodic inversion-asymmetric $Z$/Co/Pt multilayers of monatomic layer thickness for each of the three different layers [see Fig.~\ref{fig:structure}(b)] . Considering  the experience of previous DFT calculations on Co/Pt systems~\cite{Boulle:16,Yang-H:2015,Soumyanarayanan:2017,Perini:2018,Jia:18,Zimmermann:19,Jia:20}, we calculate atomistic and micromagnetic interaction parameters  characterizing the magnetic properties of the multilayers with the aim to explore in how far the Co/Pt bilayer can be turned into an interesting materials stack for AFM skyrmions in SAF by selecting appropriate nonmagnetic interlayers $Z$. We include in our exploration in total 13 different interlayers $Z$. These include the non-magnetic
4$d$ transition metals (Y--Pd), noble metals (Cu, Ag, and Au), and two of the post-transition metals, Zn and Cd [see Fig.~\ref{fig:structure}(c)]. We determine the magnetic moments, the spin stiffness, the DMI, the magnetic anisotropy, and the interlayer coupling between Co films across the structurally optimized $Z$/Pt spacers. We estimate the critical temperature using a Monte Carlo method. 

We find, independent of the spacer layer, large local magnetic moments within the Co film, with small induced moments in Pt and $Z$.  The majority of the multilayers, 10 of the 13 investigated systems exhibit critical temperatures  above room temperature, all of the 10 systems have an out-of-plane easy axis, and 6 of the 10 show an AFM interlayer coupling between the Co films forming SAF multilayers, all in all good conditions for finding AFM skyrmions. The strongest AFM interlayer coupling we found for the Rh system.

After obtaining micromagnetic parameters by microscopic DFT, we investigate larger-scale spin textures in the Co planes of the multilayers  by minimizing the materials-specific micromagnetic energy functional with respect to one- and two-dimensional textures in the sense of a multiscale model, with a focus on the formation and stability of N\'eel-type spin-spiral ground states and metastable skyrmions. We numerically solve the skyrmion profile equation using a shooting method to study the  profile and size of isolated skyrmions. At last, we develop models that allow an estimation of the skyrmion radius at finite temperature and for skyrmions in films of larger Co thicknesses based on the DFT results at zero temperature for multilayers made of monatomically thick films. 

Our findings include the following: (i) 8 of the 13 MMLs systems studied are able to form isolated metastable skyrmions at zero external magnetic field. Between the systems, the skyrmion radius can vary by two orders of magnitude. Due to the monatomic thickness of Co, all skyrmions are of atomic scale and unstable due to the granularity imposed by the underlying crystal lattice, with the exception of the Cu system, for which a skyrmion radius of 2.72 nm at zero temperature has been calculated.  (ii)  Two analytical expressions of the skyrmion radius covering the numerical results of the whole range of micromagnetic parameters.  (iii) The skyrmion radius may easily double when going from cryogenic to room temperature and (iv) tiny changes of the Co-layer thickness can change the skyrmion radius by an order of magnitude. (v) The Cu/5.8MLCo/Pt is expected to be an interesting material system for ferromagnetically and Zr/10.4MLCo/Pt, Ru/12.5MLCo/Pt, Rh/11.5MLCo/Pt for antiferromagnetically coupled skyrmion systems with a diameter of 10 nm based on the proposed model.

The paper is organized as follows: We briefly mention the structural setup and some computational details. Then, in Secs. \ref{ssec:results:structure}-D, F  we go first through the results of the different properties (structural, magnetic moments, interlayer exchange coupling, density of states, spin-stiffness, magnetic anisotropy, and DMI) obtained from \abinitio\ and report then on the critical temperature in Sec. \ref{ssec:results:Tc}. In Sec. \ref{ssec:results:NMTexture} we introduce the micromagnetic model and present  conclusions on non-trivial spin-textures and  the skyrmion radii at cryogenic and room temperature and as function of the Co thickness.  This is then followed by a conclusion. 

\section{COMPUTATIONAL DETAILS}

The DFT calculations were performed  employing the full-potential linearized augmented plane-wave (FLAPW) method as implemented in the FLEUR code~\cite{FLAPW}. All computational details and structural parameters are identical to Ref.~\@\cite{Jia:20}. The force theorem of Andersen~\cite{Mackintosh:80,Oswald:85,Liechtenstein:87} was used to calculate the spin stiffness and uniaxial magneto-crystalline anisotropy,  first-order perturbation theory in the spin-orbit interaction was used to determine the DMI~\cite{Heide:09}.  More details about the magnetic model, the extraction  of interaction parameters from DFT, and a more detailed description of the methods can be found in Refs.\@ \cite{Jia:18,Zimmermann:14}. 
The main contribution of the dipolar magnetic anisotropy to the uniaxial magnetic anisotropy was obtained by summation of classical dipole interaction terms due to local magnetic moments of the Co atoms place at lattice sites inside the Co film. Due to the wide separation of the magnetically active layers, their contributions are independent within $1\,\mu\,$eV of the type of interlayer coupling of the Co across the multilayer.

\section{Results and discussion}
\subsection{Structural properties}\label{ssec:results:structure}

\begin{figure}[t]
\centering
\includegraphics[width=85mm]{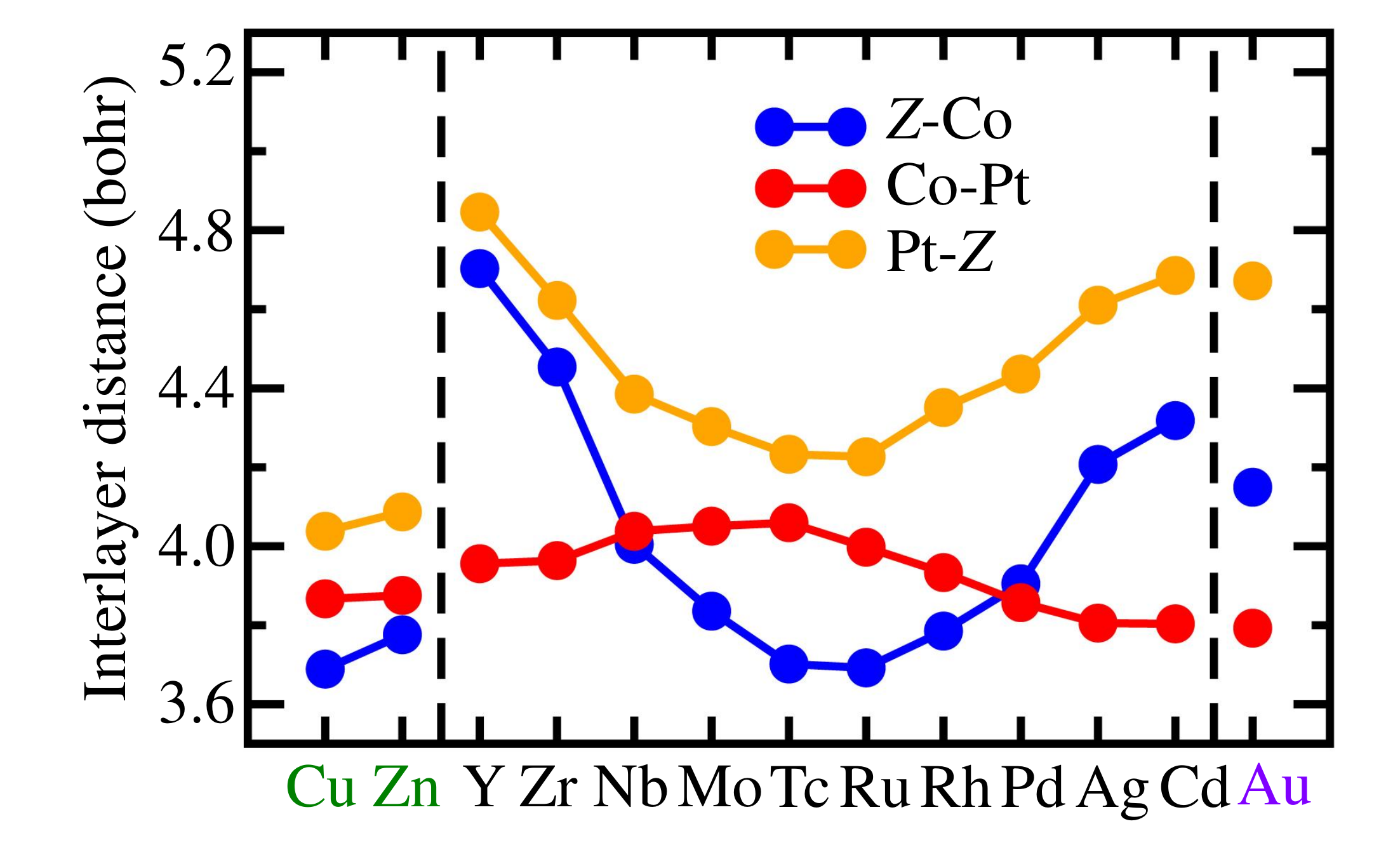}
\caption{Interlayer distances of $Z$-Co, Co-Pt, and Pt-$Z$ layers along $z$-axis direction at the FM states.}
\label{fig:distance}
\end{figure}

\begin{figure}[!t]
	\centering
	\includegraphics[width=85mm]{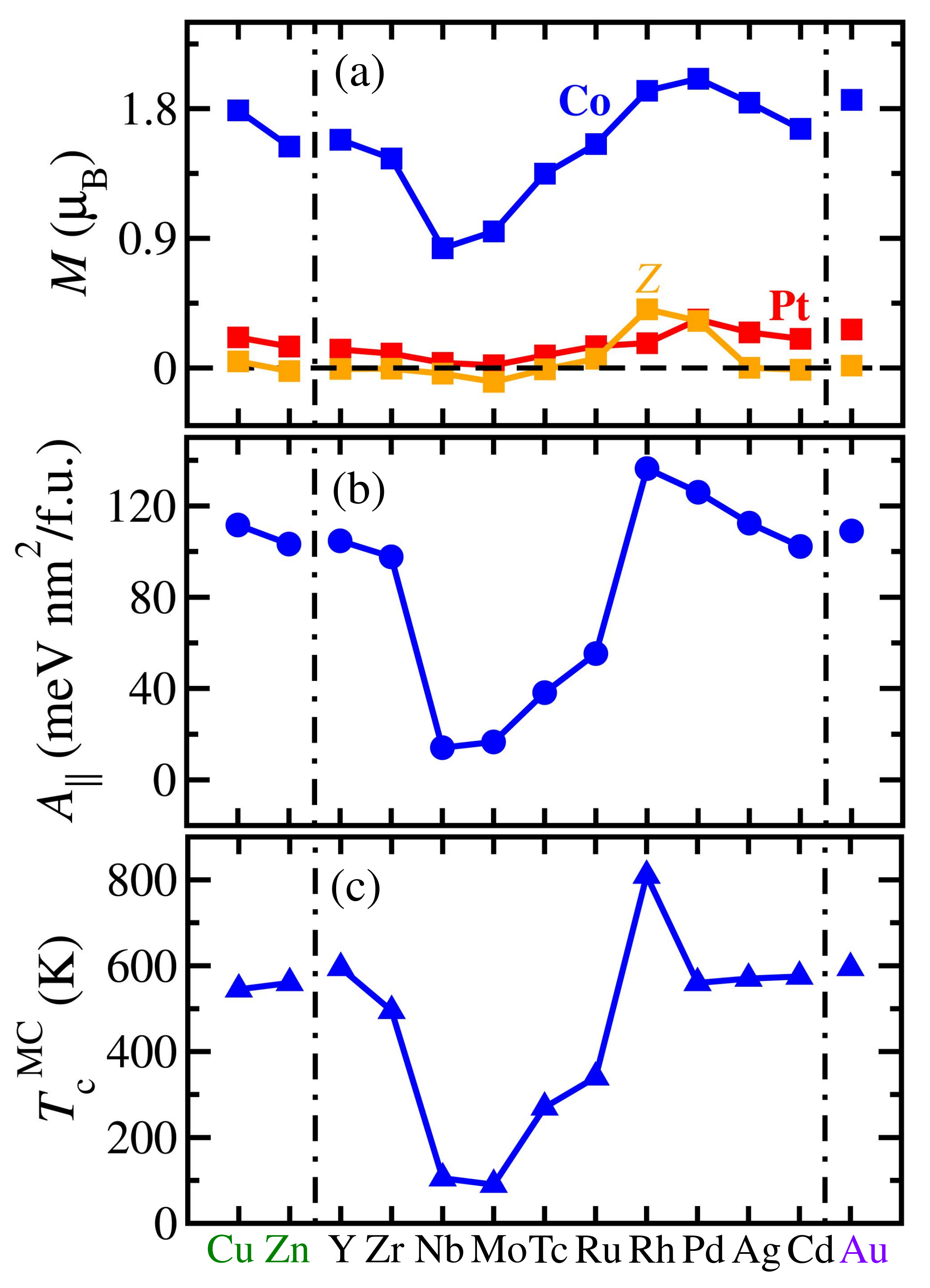}
	\caption{(a) Magnetic moments of $Z$, Co, and Pt atoms, (b) spin-stiffness constant, $A_{\|}$, extracted from spin-spiral vectors $\mathbf{q}$ along the $\overline{\Gamma \mathrm{M}}$ ($\overline{\mathrm{AL}}$) direction, and (c) Critical
		temperature for magnetic ordering of $Z$/Co/Pt ($Z=3d$:\@ Cu, Zn; 4$d$:\@ Y--Cd; 5$d$:\@ Au) MMLs at the related FM (SAF) equilibrium state.}
	\label{fig:M-A-Tc}
\end{figure}

First, we optimized the fcc-stacked crystal structure of the $Z$/Co/Pt MMLs in the FM states by fixing  the in-plane lattice constant to the experimental value of Pt(111) ($a=2.77$~\AA) and determined the optimal size of the unit cell along the $z$ axis  (the unit vector perpendicular to the multilayer plane) by optimizing the interlayer distances of $Z$-Co, Co-Pt and Pt-$Z$ layers. Since the energy scale involved in the interlayer coupling is small as compared to structural relaxations we expect that the type of coupling, FM or SAF, has basically no influence on the structural details of the MMLs. We confirmed this hypothesis analyzing one system in detail, which was Cd/Co/Pt. Figure~\ref{fig:distance} summarizes the obtained results. In the Cu/Co/Pt and Zn/Co/Pt MMLs, all atomic interlayer distances are close to each other. In the $4d$/Co/Pt MMLs, all the interlayer distances show a roughly parabolic trend, inverted for the Co-Pt compared to the $Z$-Co and Pt-$Z$ distances. An extreme point occurs at the Tc/Co/Pt system. A similar parabolic behavior has also been found in the calculated equilibrium Wigner-Seitz radii of the 4$d$ metals in Ref.\@ \cite{Methfessel:92}, which demonstrates that the size of the 4$d$ atoms not only influences the interlayer distances involving the $4d$ atom, but also the interlayer distance of Co-Pt. It should be noted that there are two intersections between the interlayer distance of Co-Pt and Co-$4d$ layers occurring at Nb/Co/Pt and Pd/Co/Pt MMLs, respectively. In other words, the Co layer is much closer to the Pt layer resulting in a stronger hybridization of the two layers in $4d$/Co/Pt ($4d=$ Y, Zr, Pd, Ag, Cd) compared to the other MMLs. This can be confirmed roughly by the spin polarization of Pt, where the induced magnetic moments of Pt in $4d$/Co/Pt ($4d=$ Y, Zr, Pd, Ag, Cd) MMLs are typically larger than in the other MMLs [see Fig.~\ref{fig:M-A-Tc}(a)]. Exceptions are the  Ru/Co/Pt and Rh/Co/Pt systems, which can be traced back to the influence of the band filling of the $4d$ electrons on the hybridization. In the Au/Co/Pt MML, the Co layer is much closer to the Pt layer than Au because of the larger atomic radius of Au.

\subsection{Interlayer exchange coupling and magnetic order in-between the magnetic layers}
\label{ssec:IEC}

\begin{figure}[!t]
	\centering
	\includegraphics[width=85mm]{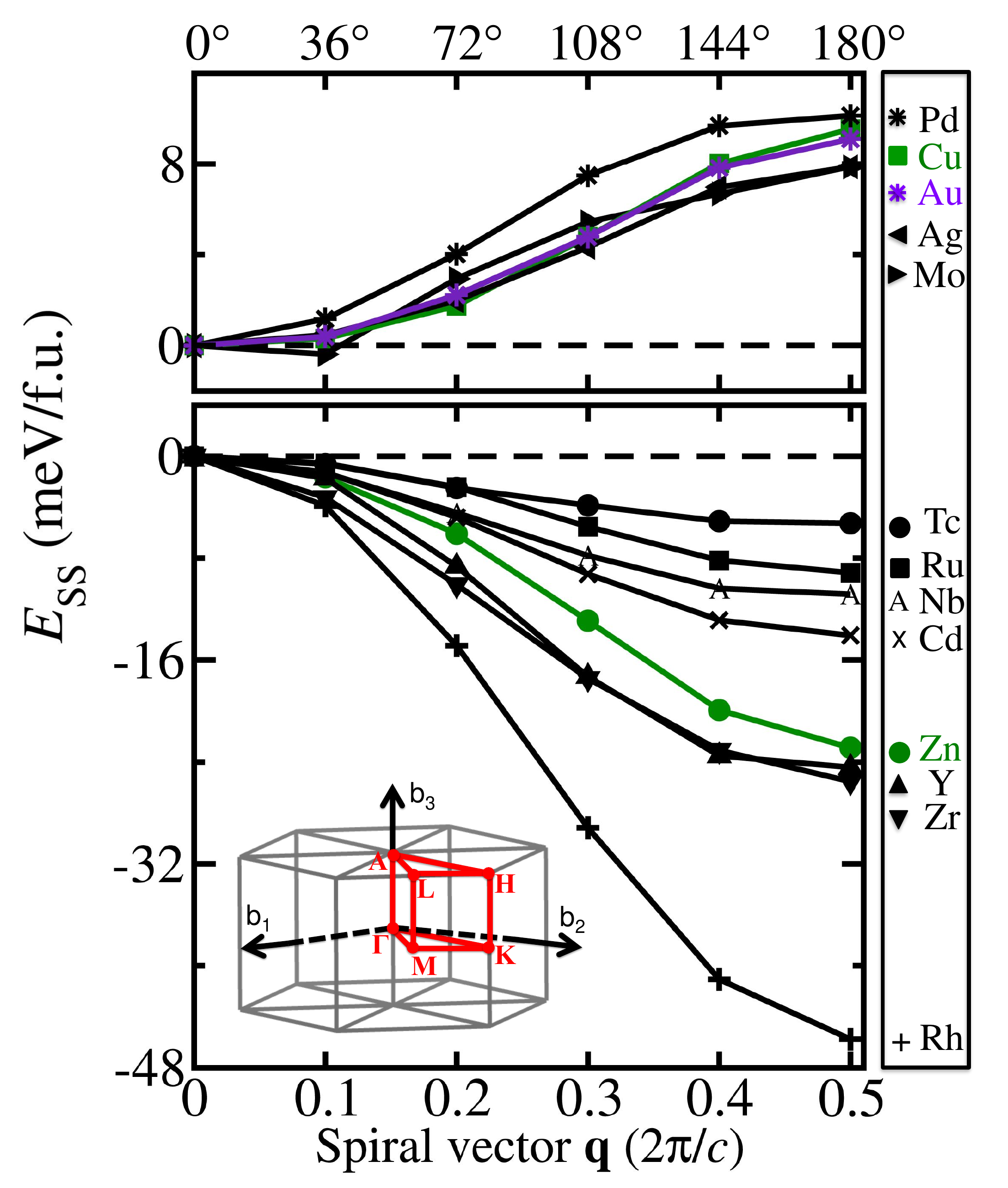}
	\caption{Energy dispersion of spin spirals in $Z$/Co/Pt ($Z=3d$:\@ Cu, Zn; 4$d$:\@ Y--Cd; 5$d$:\@ Au) MMLs for spin-spiral vectors $\mathbf{q}$ along the $\overline{\Gamma \mathrm{A}}$ direction to explore the interlayer exchange coupling. The insert shows the Brillouin zone of the hexagonal lattice. $\mathbf{b}_1$, $\mathbf{b}_2$, and $\mathbf{b}_3$ represent reciprocal lattice vectors for the chemical unit cell.}
	\label{fig:IEC}
\end{figure}

The interlayer exchange coupling between the magnetic Co layers of the MMLs was investigated based on the optimized crystal structures where a FM ordering within the planes was assumed. The energy dispersion along the high-symmetry direction $\overline{\Gamma \mathrm{A}}$ in the Brillouin zone (BZ), where the $\overline{\Gamma}$ ($\overline{\mathrm{A}}$) point represents the FM (SAF) state, is shown in Fig.~\ref{fig:IEC}. One can see that the Zn, Y, Zr, Nb, Tc, Ru, Rh, and Cd elements induce an SAF ground state between the ferromagnetic Co layers [gray color in Fig.~\ref{fig:structure}(c)], while the Cu, Mo, Pd, Ag, and Au induce a FM order [peach color in Fig.~\ref{fig:structure}(c)]. The Rh and Pd systems show the largest energy variation corresponding to the most stable AFM and FM interlayer coupling, respectively.
\begin{table}[H]
	\caption{Exchange interaction parameters ($J_n$) and resulting interlayer magnetic order for $Z$/Co/Pt ($Z=3d$:\@ Cu, Zn; 4$d$:\@ Y--Cd; 5$d$:\@ Au) MMLs. Positive (negative) $J_n$ denote (anti-)ferromagnetic interactions. SAF = synthetic antiferromagnet, FM = ferromagnet. \label{tab:iec}}
	\begin{ruledtabular}
		\begin{tabular}{rccc}
			System & Magn.\ order & $J_1$ (meV) & $J_2$ (meV) \\ \hline 
			Cu/Co/Pt &  FM & $\phantom{9}9.53$ & $-1.48\phantom{-}$  \\
			Zn/Co/Pt & SAF & $-22.88\phantom{-}$ & $1.87$ \\ \hline
			Y/Co/Pt  & SAF & $-24.40\phantom{-}$ & $-0.78\phantom{-}$ \\
			Zr/Co/Pt & SAF & $-25.53\phantom{-}$ & $-0.99\phantom{-}$  \\
			Nb/Co/Pt & SAF & $-10.85\phantom{-}$ & $-0.72\phantom{-}$  \\
			Mo/Co/Pt &  FM & $\phantom{7}7.90$ & $2.10$ \\
			Tc/Co/Pt & SAF & $-5.29\phantom{.}$ & $-0.54\phantom{-}$ \\
			Ru/Co/Pt & SAF & $-9.14\phantom{.}$ & $0.58$  \\
			Rh/Co/Pt & SAF & $-45.75\phantom{-}$ & $0.85$  \\
			Pd/Co/Pt &  FM & $10.30$ & $0.69$  \\
			Ag/Co/Pt &  FM & $\phantom{7}7.89$ & $-0.75\phantom{-}$  \\
			Cd/Co/Pt & SAF & $-14.06\phantom{-}$ & $-0.01\phantom{-}$  \\ \hline
			Au/Co/Pt &  FM & $\phantom{9}9.13$ & $-1.04\phantom{-}$ \\
		\end{tabular}
	\end{ruledtabular}
\end{table}

To quantify the interlayer exchange coupling, we mapped our results to the classical Heisenberg model 
\begin{equation}
  H=-\frac{1}{2}\sum_{i,j(i\neq{j})}{J}_{ij} \, \mathbf{e}_i \cdot \mathbf{e}_j
  \label{eq:Heisenbergmodel},
\end{equation}
where $\mathbf{e}_i$ ($\mathbf{e}_j$) is the unit vector along the moment of atom $i$ ($j$) and the magnitude of the moments is incorporated in the interaction parameters $J_{ij}$ including interactions up to the next-nearest neighbors. More details can be found in Ref.~\cite{Jia:18} where the Pd and Rh systems have been studied for different Pt and Pd/Rh thicknesses. The related parameters $J_1$ and $J_2$ are listed in Table~\ref{tab:iec}. It can be seen that the nearest-neighbor interaction dominates in most cases, with ratios $\left| J_1/J_2\right| >10$. Exceptions are the Mo, Cu, and Au systems, that all have positive $J_1$. For the latter two, the opposite signs of the two parameters even introduce some frustration to the system. On the contrary, the nearest-neighbor AFM coupling between magnetic Co layers in Cd/Co/Pt MML is extremely stable since $J_2$ is vanishingly small.

\subsection{Spin stiffness}
\label{ssec:SS}

\begin{table*}
	\caption{Magnetic moment of the Co atoms ($M_\mathrm{Co}$) and induced magnetic moment of Pt ($M_\mathrm{Pt}$) and $Z$ ($M_Z$), Critical temperature for magnetic ordering ($T^\mathrm{MC}_\mathrm{c}$), spin stiffness constant ($A_{\|}$), DMI constant ($D$) and magnetic  anisotropy constant ($K=K_\mathrm{soc}+K_\mathrm{dip}$) per formula unit (f.u.) in $Z$/Co/Pt MMLs as well as the inhomogeneity parameter $\kappa$,  domain-wall width $\ell_\mathrm{w}=(1/\pi)\sqrt{A/|K|}$, period length $\lambda_\mathrm{h(i)s}=-(\epsilon)2A/D$ of the (in)homogeneous spin spiral, and skyrmion radius, $R_\mathrm{Sk}$, calculated by the shooting method~\cite{Leonov:16,Kiselev:11,Bogdanov:94}. $D>0$ ($<0$) energy is lowered by a left (right) handed spin-configuration. $K>0$ ($<0$) indicates out-of-plane (in-plane) easy  axis. $\lambda_\mathrm{hs}=\infty$ stands for the ferromagnetic ground-state of the trilayer unit $Z$/Co/Pt and the sign of $\lambda_\mathrm{h(i)s}$ indicates  the rotational sense of the spiral ($\lambda_\mathrm{h(i)s}<0$ corresponds to counter-clockwise rotation). The magnetic parameters given in atomistic units can be translated to SI units as following: Magnetization density  $M_\mathrm{s}$ [MA/m] $=0.70\,M_\mathrm{s}$ [$\mu_\mathrm{B}$], $A_{\|}$ [pJ/m]$=12.06\,\frac{A_{\|}}{4\pi^2}$ [meV\,nm$^2$/f.u.], $D$ [mJ/m$^2$] $= 12.06\,\frac{D}{2\pi}$ [meV\,nm/f.u.], $K$ [MJ/m$^3$] $=12.06\,K$ [meV/f.u.]. We take $t_\mathrm{Co}=0.2$~nm as the thickness of the Co monolayer, which corresponds to a Co volume of $0.0133\,$nm$^3$.
	}

	\label{tab:table}
	\begin{ruledtabular}
		\begin{tabular}{crccccccccccc}
			& System & \multicolumn{1}{c}{$M_\mathrm{Co}$} & \multicolumn{1}{c}{$M_\mathrm{Pt}$} & \multicolumn{1}{c}{$M_Z$} &\multicolumn{1}{c}{$T^\mathrm{MC}_\mathrm{c}$}& \multicolumn{1}{c}{$A_{\|}$} & \multicolumn{1}{c}{$D$} & \multicolumn{1}{c}{$K$} & $\kappa$ & $\ell_\mathrm{w}$ & $\lambda_\mathrm{h(i)s}$ & $R_\mathrm{Sk}$  \\
			&     {}  & ($\mu_\mathrm{B}$)                  & ($\mu_\mathrm{B}$)                  & ($\mu_\mathrm{B}$)                  & (K) &  \multirow{2}{*}{\large($\frac{\mathrm{meV\,nm}^2}{\mathrm{f.u.}}$)}  & \multirow{2}{*}{\large($\frac{\mathrm{meV\,nm}}{\mathrm{f.u.}}$)}         & 	\multirow{2}{*}{\large($\frac{\mathrm{meV}}{\mathrm{f.u.}}$)} & {}       & (nm) & (nm) & (nm) 
			\\ & & & & & & & & & & & & \\   \hline
			\multirow{2}{*}{$3d$} & Cu/Co/Pt & 1.79 &0.21&0.05&545& 112 & 6.66 &0.46& 1.88 &4.97 & $\infty$ & 2.72 \\
			& Zn/Co/Pt & 1.54 &0.15&$-0.02\phantom{-}$&560& 103 & 7.54 &1.53& 4.49 & 2.61 & $\infty$ & 0.44 \\ \hline
			\multirow{10}{*}{$4d$} & Y/Co/Pt & 1.58 &0.13&$-0.01\phantom{-}$&595& 105 & 2.70 &1.05& $24.52\phantom{4}$ & 3.18 & $\infty$ & 0.10 \\
			& Zr/Co/Pt & 1.45 &0.10&$-0.00\phantom{-}$&495& $\phantom{5}98$ & $-0.16\phantom{-}$ &0.42& $2606.49\phantom{111}$ & 4.86 & $\infty$ & 0.01 \\
			& Nb/Co/Pt & 0.83 &0.04&$-0.04\phantom{-}$&105& $\phantom{5}14$ & $-0.35\phantom{-}$ &$-0.06\phantom{-}$& $11.12\phantom{4}$ & 4.86 & $\infty$ & \\
			& Mo/Co/Pt & 0.95 &0.02&$-0.10\phantom{-}$&\phantom{0}90& $\phantom{5}17$ & $-1.12\phantom{-}$ &$-0.09\phantom{-}$& 1.98 & 4.37 & $\infty$ & \\
			& Tc/Co/Pt & 1.35 &0.09&$-0.01\phantom{-}$&270& $\phantom{5}38$ & $-2.80\phantom{-}$ &$-0.17\phantom{-}$& 1.34 & 4.76 & $\infty$ & \\
			& Ru/Co/Pt & 1.55 &0.15&0.06&340& $\phantom{5}55$ & 0.43 &0.57& $274.87\phantom{66}$ & 3.13 &  $\infty$ & 0.02 \\
			& Rh/Co/Pt & 1.92 &0.17&0.41&810& 136 & $-1.62\phantom{-}$ &0.79& $66.37\phantom{4}$ & 4.18 & $\infty$ & 0.06 \\
			& Pd/Co/Pt & 2.01 &0.34&0.33&560& 126 & 4.39 &1.51& $16.00\phantom{4}$ & 2.91 & $\infty$ & 0.13 \\
			& Ag/Co/Pt & 1.84 &0.25&0.00&570& 112 & 6.48 &0.06& 0.26 &  & $-34.57$ & \\
			& Cd/Co/Pt & 1.66 &0.20&$-0.01\phantom{-}$&575& 102 & $11.44\phantom{6}$ &0.74& 0.93 &  & $-25.03$ & \\ \hline
			\multirow{1}{*}{$5d$} & Au/Co/Pt & 1.86 &0.27&0.02&595& 109 & 3.15 &0.60& $10.69\phantom{4}$ & 4.29 & $\infty$ & 0.29\\
		\end{tabular}
	\end{ruledtabular}
\end{table*}

The in-plane spin stiffness, $A_{\|}$, was extracted from spin-spiral calculations along the $\overline{\Gamma \mathrm{M}}$ ($\overline{\mathrm{AL}}$) direction in the BZ around the respective lowest-energy state, \ie the FM (SAF) state, as determined by the interlayer exchange coupling [see Fig.~\ref{fig:M-A-Tc}(b)]. $A_{\|}/(4\pi^2)$ was obtained by quadratic fits of the energy dispersion $E(\mathbf{q})$ in the range $\mathbf{q} < 0.1 (2\pi / a)$, \eg $E(\mathbf{q})\simeq \frac{A_{\|}}{4\pi^2}\vc{q}^2$ in the vicinity of the $\Gamma$-point. For a few systems we analyzed the effect of the type of interlayer exchange coupling on the in-plane spin-stiffness. For the systems studied, $A_{\|}$ various  between 0 to 4\% when changing the type of interlayer coupling. 

 The variation of the spin stiffness across the transition metal (TM) elements around the respective lowest-energy states can be found in Fig.~\ref{fig:M-A-Tc}(b). It is obvious that the spin-stiffness constant is largest for the Rh/Co/Pt system and decreases gradually for the other $4d$/Co/Pt MMLs for chemical elements on either the left or right side of Rh in the 4$d$ TM row of the periodic table. A similar trend has been reported by Parkin~\cite{Parkin:91}. He observed a larger in-plane magnetic exchange coupling for a Co/Rh multilayer compared to the ones in Co/$Z$ ($Z=$ Cu, Ru, Mo, Nb). The values of the spin stiffness constant $A_{\|}$ are listed in Table~\ref{tab:table}.
 
 The spin stiffness depends mainly on the non-relativistic exchange interactions and represents the change in the ground-state energy caused by the twist of the local moments along the plane of the MML. The correlation with the variation of the Co magnetic moments can be seen from from Fig.~\ref{fig:M-A-Tc}(a): Although the curve's maximum occurs at the Pd/Co/Pt system, the Co magnetic moment and the spin stiffness follow nearly the same trend for these MMLs. The Co atom has the largest magnetic moment in the Pd/Co/Pt system and the moments gradually decrease for the other $4d$/Co/Pt MMLs with the 4$d$ TM elements on both side of Pd. Such a decrease of the magnetic moment of Co atom from Rh to Ru was also reported by S. Zoll \etal \cite{Zoll:97}:\@ Compared to bulk Co, a strong reduction of the Co magnetization was observed in the Co/Ru multilayer system experimentally, while no variation was detected in the Co/Rh case. Due to the strong orbital hybridizations across the interface, magnetic moments are induced in the Pt atoms. They are ferromagnetically coupled to the adjacent Co moment and their size is roughly proportional to the Co moment. The other TM atoms have a rather weak magnetic polarization with the exception of Rh and Pd.

\begin{figure}[!t]
\centering
\includegraphics[width=80mm]{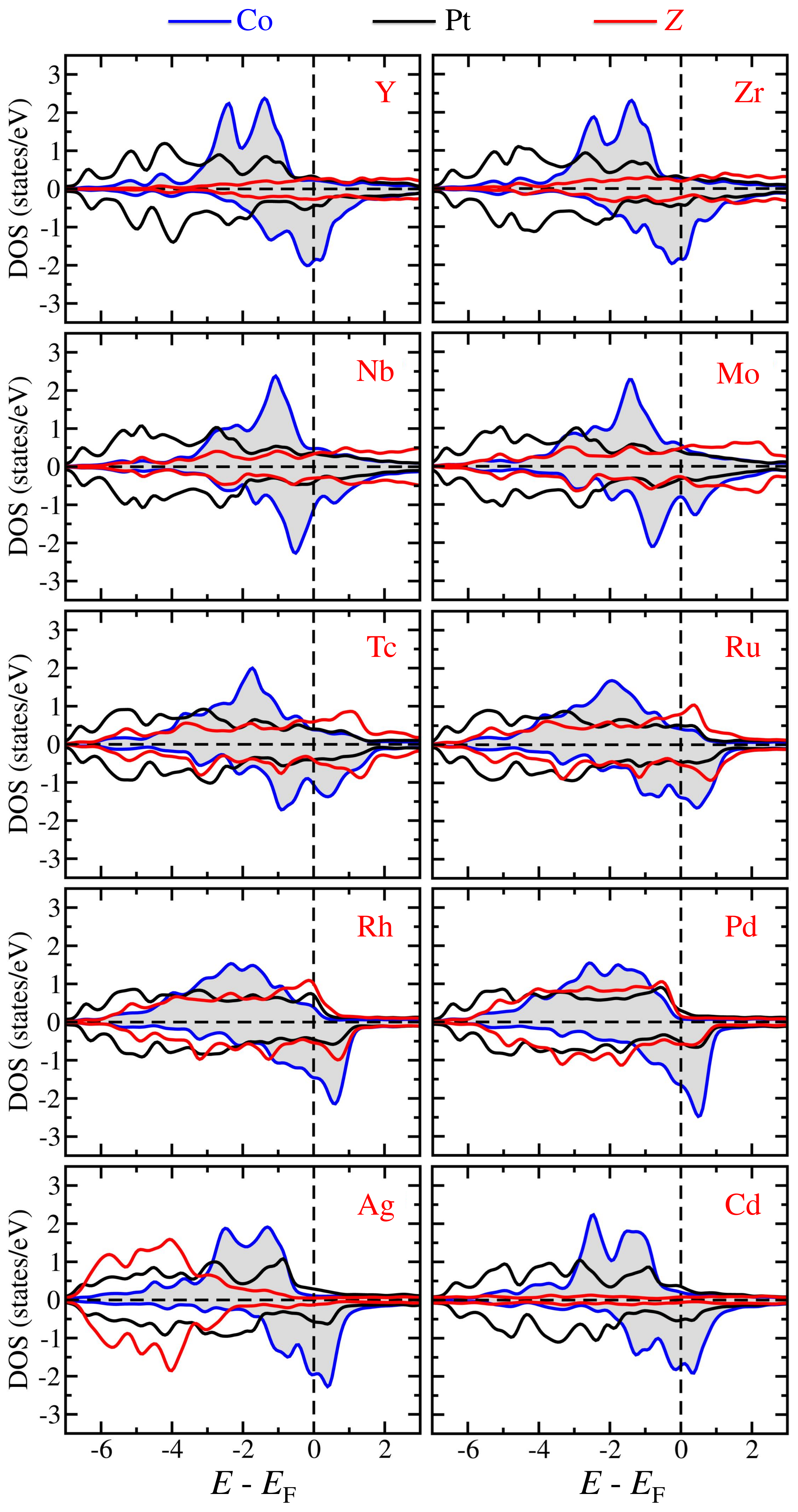}
\caption{Projected DOS of 4$d$, Co and Pt layers in majority and minority spin channels calculated for $4d$/Co/Pt ($4d=$ Y, Zr, Nb, Mo, Tc, Ru, Rh, Pd, Ag, Cd) MMLs at their equilibrium states.}
\label{fig:dos}
\end{figure}

The observed trend of the spin stiffness and magnetic moments is correlated to the TM $d$-$d$ orbital hybridizations around the Fermi level and the band filling of the $4d$-spacer layers. To illustrate this, we calculated the local density of states (DOS) of the different atomic layers exemplary for the $4d$/Co/Pt MMLs [see Fig.~\ref{fig:dos}]. Based on our results, the outermost $d$-orbital electrons of each TM atomic layer give a major contribution to the total DOS around the Fermi level, while the $s$-orbital and $p$-orbital electrons just have little effect (data not shown). Starting on the left side of the periodic table, \ie those with least number of $d$ electrons, Y and Zr, one can see a only small DOS coming from the $4d$ elements around the Fermi level [see Fig.~\ref{fig:dos}]. This indicates a very weak 3$d$-4$d$ hybridization with the Co atoms and the overall DOS around the Fermi level is mainly determined by the hybridization between the Co and Pt atoms. Going to Nb and Mo, it can be seen that the DOS around the Fermi level is increasing due to the increasing filling of the 4$d$ orbital. This reduces the spin-splitting of Co resulting in a sharp reduction of its magnetic moment [see Fig.~\ref{fig:M-A-Tc}(a)]. Increasing the $4d$-orbital filling further, the number of unoccupied states above the Fermi level decreases in both spin channels (Y$\rightarrow$Pd). Tc, Ru, Rh, and Pd atoms show a sizable DOS in both majority and minority spin channels around the Fermi level. In particular, Rh being isoelectronic to Co, and Pd being isoelectronic to Pt, have the strongest 3$d$-4$d$-5$d$ orbital hybridizations due to their similar electronic structures with Co and Pt, respectively. The largest magnetic moment of Co and induced magnetic moment of Pt are obtained in the Pd/Co/Pt MML due to the strong spin polarization at the Fermi level. Rh gets the largest induced magnetic moment in comparison to other $4d$ elements. Finally, the electronic states of Ag and Cd are shifted towards lower energy and their DOS is vanishingly small at the Fermi level. Thus, the filling of external orbitals decreases the hybridizations of the 4$d$ ($4d=$ Ag and Cd) with the Co atoms significantly. However, the strong 3$d$-5$d$ orbital hybridizations still exist between the Co and Pt atomic layers indicated by the sharp peaks below the Fermi level ($-3$ to 0 eV) in the majority-spin channel, whereas the appearance of peaks below the Fermi level ($-2$ to $0$ eV) in the minority spin channel reduces the spin splitting. Therefore, the magnetic moments of Co decrease slightly in Ag/Co/Pt and Cd/Co/Pt MMLs.

\subsection{Magnetic anisotropy}

\begin{table}
	\caption{Magnetic anisotropy constant ($K$), magnetocrystalline anisotropy induced by the spin-orbit interaction ($K_\mathrm{soc}$) and magnetic dipole-dipole interaction ($K_\mathrm{dip}$) of $Z$/Co/Pt ($Z=3d$:\@ Cu, Zn; 4$d$:\@ Y--Cd; 5$d$:\@ Au) MMLs in their equilibrium states. The negative (positive) sign indicates a preferred in-plane (out-of-plane) orientation of the magnetization.}
	\label{tab:anisotropy-K}
	\begin{ruledtabular}
		\begin{tabular}{rccc}
			System &\multicolumn{1}{c}{$K$}& \multicolumn{1}{c}{$K_\mathrm{soc}$} & \multicolumn{1}{c}{$K_\mathrm{dip}$} \\
			{}  & ({meV}/{f.u.}) &     (meV/{f.u.})    & ($\mu$eV/{f.u.}) \\  \hline
			Cu/Co/Pt &0.46& 0.53 & $-66.28$  \\
			Zn/Co/Pt &1.53& 1.58 & $-48.97$ \\ \hline
			Y/Co/Pt  &1.05& 1.10 & $-52.13$ \\
			Zr/Co/Pt &0.42& 0.46 & $-44.04$  \\
			Nb/Co/Pt &$-0.06\phantom{-}$& $-0.05\phantom{-}$ & $-14.25$  \\
			Mo/Co/Pt &$-0.09\phantom{-}$& $-0.07\phantom{-}$ & $-18.75$ \\
			Tc/Co/Pt &$-0.17\phantom{-}$& $-0.13\phantom{-}$ & $-37.91$ \\
			Ru/Co/Pt &0.57& 0.62 & $-49.77$  \\
			Rh/Co/Pt &0.79& 0.87 & $-75.78$  \\
			Pd/Co/Pt &1.51& 1.59 & $-83.96$  \\
			Ag/Co/Pt &0.06& 0.13 & $-70.29$  \\
			Cd/Co/Pt &0.74& 0.80 & $-57.26$  \\ \hline
			Au/Co/Pt &0.60& 0.67 & $-72.04$ \\
		\end{tabular}
	\end{ruledtabular}
\end{table}   

Due to the two-dimensional character of metallic multilayers and the monatomic thickness of the ferromagnetic Co layer, their magnetic anisotropy is mainly determined by the interface anisotropy. The interface anisotropy contains contributions from the magnetic dipolar interaction and the spin-orbit interaction. We approximate our analysis by considering just uniaxial anisotropy with its distinguished magnetization axis pointing along the $z$ direction. Thus, we calculate the magnetic anisotropy coefficient $K$ from the energy difference between energies of two different magnetization directions, the one  pointing in plane, along the $x$ axis, and one pointing out of plane, along the $z$ axis. 

Table~\ref{tab:anisotropy-K} lists the total magnetic anisotropy coefficient $K=K_\mathrm{soc}+K_\mathrm{dip}$ and information about both contributions separately. All systems have been considered in their magnetic ground state according to Table~\ref{tab:iec}. 
It is clear that the main contribution originates from the spin-orbit interaction, while the one from magnetic dipolar interactions is one or two orders of magnitude smaller. It is often enlightening  to translate the physical origin motivated separation of the magnetic anisotropy into a micromagnetic one $K=K_\mathrm{soc}^\mathrm{MCA}+K_\mathrm{dip}^\mathrm{MCA} + K_\mathrm{dip}^\mathrm{shape}$ where we separate according the magnetocrystalline anisotropy (MCA) and the shape anisotropy. Both terms, $K_\mathrm{soc}$ and $K_\mathrm{dip}$,  contribute to magnetocrystalline anisotropy, $K^\mathrm{MCA}$,  $K_\mathrm{soc}$ contributes to 100\% and $K_\mathrm{dip}$ to a tiny fraction. The dipolar part to the magnetic anisotropy contributes mostly to the shape anisotropy constant, $K_\mathrm{dip}\simeq K_\mathrm{dip}^\mathrm{shape}= K_\mathrm{shape}=-\pi\alpha^2M_\mathrm{Co}^2/V$ in atomic Rydberg units/f.u.\ of a perfectly flat Co film of infinite extension with the local magnetic moment $M_\mathrm{Co}$ in the effective micromagnetic volume $V$ of the Co film given of 114.278~a.u.$^3$ per Co atoms corresponding to a thickness of the Co film of 2.55~\AA, and $\alpha$ is the fine-structure constant. The deviation of $K_\mathrm{shape}$ from $K_\mathrm{dip}$ we associate with the dipolar contribution to the magnetocrystalline anisotropy which amounts in average to about ${0.27}\,\mu$eV. The fact that $K_\mathrm{dip}$ is basically absorbed by $K_\mathrm{shape}$ is consistent with our analysis that the $K_\mathrm{dip}$ is practically independent of the type of magnetic coupling of the Co layers across the interlayers.  
The sign of $K_\mathrm{shape}$ or  $K_\mathrm{dip}$, respectively, remains negative as the energy of the shape anisotropy can be minimized for an in-plane orientation of the magnetization. 

This is  consistent with other thin-film systems \cite{Draaisma:88,Bruno:89}. For example, a free-standing hexagonal Co monolayer was predicted to have an in-plane magnetic anisotropy~\cite{Daalderop:94}. However, a combination of Co and non-magnetic metals, such as Co/Pd(111) and Co/Pt(111), shows a strong perpendicular magnetic anisotropy (PMA)~\cite{Hashimoto:89,Nakajima:98} and it was shown that the main contribution originates from the non-magnetic atoms at the magnetic-nonmagnetic interface~\cite{Zimmermann:14,Khmelevskyi:11}. In our work, all systems show a PMA, except for the Nb, Mo, and Tc systems. Unfortunately, it is difficult to obtain separately the contribution of each atomic layer  to the magnetocrystalline anisotropy due to the strong hybridization across the interfaces. Nevertheless, the results clearly demonstrate the important role of the different nonmagnetic spacer layers in the formation of the magnetic anisotropy of MMLs. Therefore, it is possible to tailor the magnetic anisotropy of metallic multilayers by choosing appropriate materials.

\subsection{Critical temperature}
\label{ssec:results:Tc}

The magnetic ordering temperature,  $T_\mathrm{c}$, of the system-specific classical Heisenberg model~\eqref{eq:Heisenbergmodel} extended by a term including the magnetic anisotropy, $K$, is determined by means of the Monte Carlo (MC) method applying the Metropolis algorithm for the importance sampling of the phase space as implemented is the SPIRIT code~\cite{spirit,muller:19}. This method provides unbiased critical temperatures for FM and SAF multilayers alike. With ``unbiased'' we refer to the fact that in this MML systems the magnetic interactions are very anisotropic, meaning a strong ferromagnetic exchange interaction in the Co layer meets a relatively weak ferromagnetic or antiferromagnetic one between the Co layers. In this case the physically transparent three-dimensional mean-field results can differ substantially from the numerically precise Monte Carlo results. We found that in average the mean field result for the ferromagnetically coupled layers calculated by the formula $k_\mathrm{B}T^\mathrm{MFA}_\mathrm{C}=\frac{2}{3}\sum_{i,j(i\neq{j})}{J}_{ij}$ is about  59\% larger than the MC data. For the antiferromagnetically ordered multilayers with a ground state at $\mathbf{q}=\mathbf{Q}$ the mean field result obtained by $k_\mathrm{B} T^\mathrm{MFA}_\mathrm{N} = \frac{2}{3} J(\mathbf Q)$ are systematically too small.

We consider interactions up to the 15th-neighbor shell, including also interactions between the Co layers. The interaction parameters are extracted using the computationally efficient magnetic force theorem \cite{Mackintosh:80,Oswald:85,Liechtenstein:87}. The starting point of these calculations is the collinear state of lowest energy determined self-consistently.

The results are listed in Table~\ref{tab:table}. It can be seen that the critical temperatures are ranging from $90$ to $810\,$K for these MMLs. The Curie temperature of an ideal monolayer Co on Pt(111) is roughly 500-$623\,$K, which has been reported both in experiment~\cite{Shern:99} and theory~\cite{Zimmermann:19}. Thus, our results suggest a similar $T_\mathrm{c}$ in the multilayers containing Cu, Pd, Ag, Au, Cd, Zn, Y and Zr, in the case of Rh it is even enhanced. For multilayers containing Mo, Tc, Ru and Nb we find a reduced ordering temperature, correlated with the small spin stiffness $A_{\|}$. As can be seen from the comparison between Ag and Mo, the interlayer coupling, $J_1$, (Table~\ref{tab:iec}) is large enough to stabilize long range magnetic order as long as the intraplane interactions are strong. The correlation between $T_\mathrm{c}$ and $A_{\|}$ can be seen clearly in Fig.~\ref{fig:M-A-Tc} comparing panels (b) and (c). Some enhancement of $T_\mathrm{c}$ in the case of Rh/Co/Pt can be traced back to the large $J_1$ in this case. The magnetic anisotropy has a minor stabilizing effect, e.g.\ in the case of Ru/Co/Pt it leads to 6\% enhancement of $T_\mathrm{c}$ (here $K/J_1$ is 0.06).

\subsection{Dzyaloshinskii-Moriya interaction}
\label{ssec:results:DMI}

Using the scalar-relativistic approximation and treating SOC within first-order perturbation theory, we calculated the DMI energies of spin spirals for the $Z$/Co/Pt MMLs. The layer-resolved DMI energy was additionally obtained by decomposing the SOC operator into contributions of each atom in the unit cell separately. The total DMI strength of each $Z$/Co/Pt MML and the contributions of different atomic layers were extracted through a cubic fit to the DMI energy $E_\mathrm{DMI}(\mathbf{q}) = \frac{D}{2\pi} \vert\mathbf{q}-\mathbf{q}_0\vert + C \vert\mathbf{q}-\mathbf{q}_0\vert^3$ in the vicinity of the ground state $\mathbf{q}_0$ where the linear part dominates, \ie  $\vert\mathbf{q}-\mathbf{q}_0\vert < 0.1\, (2\pi / a)$. The calculations were performed for spin-spiral vectors $\mathbf{q}$ along the $\overline{\Gamma \mathrm{M}}$ or $\overline{\mathrm{AL}}$ high-symmetry line of the BZ,
depending on the interlayer interaction (Table~\ref{tab:iec}).

The total DMI strengths are listed in Table~\ref{tab:table}. First, we can see that the DMI can be tuned in a large range of $-2.8$ to $11.4\,$meV\,nm/f.u.\ by varying the non-magnetic spacer layers in the MMLs. For the convenience of analysis, we divide the $Z$/Co/Pt MMLs into two groups:\@ (I) the MMLs with $Z$ elements on the right side of Co in the periodic table (\ie Cu, Zn, Pd, Ag, Cd, and Au) with fully occupied $d$ orbitals; (II) the MMLs with $Z$ on the left side or same column of Co (Y, Zr, Nb, Mo, Tc, Ru, and Rh). 

\begin{table}[H]
 \caption{Layer-resolved DMI contributions of the different atomic layers of $Z$/Co/Pt ($Z=3d$:\@ Cu, Zn; 4$d$:\@ Y--Cd; 5$d$:\@ Au) MMLs in the respective equilibrium state.}
 \label{tab:DMI}
 \begin{ruledtabular}
  \begin{tabular}{rccc}
System & \multicolumn{1}{c}{$D_\mathrm{Pt}$} & \multicolumn{1}{c}{$D_\mathrm{Co}$} & \multicolumn{1}{c}{$D_{Z}$} \\
    {}  & (meV nm/{f.u.})      &     (meV nm/{f.u.})    & (meV nm/{f.u.}) \\  \hline
 Cu/Co/Pt & \phantom{ 2}7.33 & $-0.20\phantom{-}$ & $-0.46\phantom{-}$  \\
 Zn/Co/Pt & \phantom{ 2}8.95 & $-1.20\phantom{-}$ & $-0.21\phantom{-}$ \\ \hline
 Y/Co/Pt  & \phantom{ 2}3.11 & $-0.20\phantom{-}$ & $-0.20\phantom{-}$ \\
 Zr/Co/Pt & $-0.10$ & 0.50 & $-0.57\phantom{-}$  \\
 Nb/Co/Pt & $-0.61$ & 0.46 & $-0.20\phantom{-}$  \\
 Mo/Co/Pt & $-1.95$ & 0.17 & 0.65 \\
 Tc/Co/Pt & $-3.85$ & $-1.57\phantom{-}$ & 2.63 \\
 Ru/Co/Pt & \phantom{ 2}1.51 & $-1.31\phantom{-}$ & 0.23  \\
 Rh/Co/Pt & $-2.22$ & 0.43 & 0.17  \\
 Pd/Co/Pt & \phantom{ 2}4.99  & $-0.35\phantom{-}$ & $-0.26\phantom{-}$  \\
 Ag/Co/Pt &  \phantom{ 2}8.20 & $-0.74\phantom{-}$ & $-0.98\phantom{-}$  \\
 Cd/Co/Pt & \phantom{-}12.19 & $-0.42\phantom{-}$ & $-0.33\phantom{-}$  \\ \hline
 Au/Co/Pt & \phantom{ 2}8.70 & $-0.60\phantom{-}$ & $-4.94\phantom{-}$ \\
  \end{tabular}
 \end{ruledtabular}
\end{table}

In comparison to the group (II) MMLs, a significantly larger total DMI was obtained in the group (I) MMLs. In particular, the Cd/Co/Pt system exhibits a very strong DMI of $11.44\,$meV nm/f.u., which increases by 61\% in comparison to the single Co/Pt interface ($7.1\,$meV nm/f.u., taken from Freimuth \etal~\cite{Freimuth:14}). We further found that the Pt layer makes an essential contribution of more than $80$\% to the total DMI in these group (I) MMLs, while Co and all non-magnetic spacer layers in this group, except Au, have just weak contributions of opposite sign (see Table~\ref{tab:DMI}). We could attribute this to the short interlayer distance between the Co and Pt atoms and the fully occupied $d$ orbitals of the $Z$ atoms (see Secs.~\ref{ssec:results:structure} and \ref{ssec:SS}) in these MMLs, which facilitates a stronger hybridization and DMI at the Co-Pt interface. However, Au/Co/Pt MML is an exception within group (I). This is because of a larger SOC contribution ($-35$\%) with opposite sign compared to the Pt layer contribution (61\%), leading to a compensation and a decrease of the total DMI magnitude. In addition to the crystal and electronic structures, another possible reason for the large DMI of the group (I) MMLs is the charge transfer and the respective potential gradient \cite{Jia:18,Belabbes:16}. This can influence the size of the DMI as an additional factor in two-dimensional systems. Considering the CoPt$Z$ trilayer in the MMLs and taking into account that the non-magnetic metallic layer atoms $Z$ of group (I) have a filled $d$ shell similar to Pt, it is reasonable that the largest charge and potential gradients occur at the CoPt interface. In particular, Pd is isoelectronic to Pt and the maxima of the DOS for $Z$ ($Z=3d$:\@ Cu, Zn; $4d$:\@ Ag, Cd; $5d$:\@ Au) are far away from the Fermi level [see Fig.~\ref{fig:dos}]. 

However, the situation is different for the group (II) MMLs of composition $4d$/Co/Pt ($4d=$ Y, Zr, Nb, Mo, Tc, Ru, Rh). They show significantly larger Co-Pt interlayer distances and less valence electrons in the non-magnetic metallic layers $Z$ than in the Co layer, resulting in charge and potential gradients occurring at the Pt-$Z$ or $Z$-Co instead of the Co-Pt interface. This reduces the magnitudes of the DMI significantly and even changes the chirality of the MMLs (see Table~\ref{tab:table}). It should be noted that the Pt layer still shows a dominant contribution to the total DMI in most of the group (II) MMLs.

As seen in Sec.~\ref{ssec:IEC}, the non-magnetic elements Zn, Y, Zr, Nb, Tc, Ru, Rh, and Cd induce an AFM coupling between the magnetic Co layers. Therefore, the DMI shown in Tables~\ref{tab:table} and \ref{tab:DMI} was calculated at their equilibrium SAF states. For comparison we also calculated the DMI of these $Z$/Co/Pt MMLs in their FM states. Although the magnitudes of total DMI for the FM state vary slightly compared to the SAF state, the signs of DMI are consistent between FM and SAF states in most of the MMLs, except for the Rh/Co/Pt MML. We found that the Pt layers still dominate the total DMI. A more detailed analysis on the sign and magnitude of DMI modified by the non-magnetic $Z$ elements can be found in Ref.~\cite{Jia:20}, where it is demonstrated that the electric and magnetic dipole moments of the Pt atoms show a strong correlation with the DMI at the FM states.

\subsection{Nontrivial magnetic textures}
\label{ssec:results:NMTexture}

\subsubsection{Micromagnetic model}
\label{ssec:MModel}
The starting point of our considerations had been the Co/Pt interface which supports a ferromagnetic out-of-plane ground-state magnetic structure. If we take the micromagnetic parameters as calculated for the $Z$/Co/Pt multilayer, the question arises to what extent the presence of the second interface and the change in effective interactions due to the embedding of the trilayer system in a multilayer changes our basic assumptions about the ground state.  

For this purpose we take here a multiscale approach and analyze the magnetic texture in an effectively single Co film on the basis of the so-called \abinitio\ micromagnetic continuum theory described by the energy  functional for structure inversion asymmetric interfaces 

\begin{eqnarray}
\label{eq:1}
E[\mathbf{m}] &= {\displaystyle \int} \bigg [&\phantom{+}\frac{A_{\|}}{4\pi^2}(\nabla \mathbf{m})^2\nonumber\\ 
&&+\frac{D}{2\pi}[(\mathbf{m}\cdot\nabla)m_z-m_z(\nabla\cdot \mathbf{m})]\nonumber\\
&&-Km^{2}_{z}-Bm_z\bigg ]\diff{\vc{r}_{\scriptscriptstyle \Vert}}\, ,  
\end{eqnarray}
where $\vc{m}(\vc{r}_{\scriptscriptstyle \Vert})=\vc{M}(\vc{r}_{\scriptscriptstyle \Vert})/M_0$ is a continuous unit magnetization field given in Cartesian coordinates, $B$ is the external magnetic field applied perpendicular to the Co film plane with the micromagnetic parameters obtained for the different multilayers by density functional theory and taken from Table~\ref{tab:table}, but ignoring any additional interlayer exchange coupling between the Co films. The latter assumption means to a good approximation, we assume the same behavior in all Co films of the multilayer, which determines finally the magnetic state of the MMLs. We assume Co films limited to thicknesses such that the magnetostatic energy not explicitly included in \eqref{eq:1} does not introduce a three-dimensional spatial inhomogeneity and we can safely assume that the magnetization density is independent of the coordinate normal to the film direction, \ie $z$ direction, $\vc{m}(\vc{r}_{\scriptscriptstyle \Vert},z)\approx \vc{m}(\vc{r}_{\scriptscriptstyle \Vert})$. Under this assumption the magnetostatic energy of a single, infinitely extended film can be separated into two contributions, the previously discussed shape anisotropy that depends on the out-of-plane magnetization $m_z^2$ and has the same functional form as the  magnetic anisotropy and is thus included in the micromagnetic parameter $K$, and the second one the film-charge term, $\sigma(\vc{r}) = \nabla \cdot \vc{m}_\parallel$, which depends on the in-plane magnetization. The latter is not included but briefly discussed in section~\ref{ssec:Rsk-10}.

We discuss  the stability and in particular the experimentally measurable size of the magnetic textures. We analyze the magnetic profile minimizing the micromagnetic energy functional \eqref{eq:1} in one and two dimensions in terms of the  effective materials-specific parameter  $\kappa=\left(\frac{4}{\pi}\right)^2 \frac{A_{\|} K}{D^2}$, determined by the in-plane spin stiffness ($A_{\|}$), magnetic anisotropy ($K$), and DM interaction strength ($D$)~\cite{Heide:11,Dzyaloshinskii:65,Izyumov:84,Jia:18}. To simplify expressing the solutions derived from \eqref{eq:1} it is convenient to introduce typical micromagnetic length scales whose physical meaning becomes transparent below. Among those are the length scale of a DMI driven homogeneous spin-spiral,  $\ell_\mathrm{hs}=2{A_{\|}}/\abs{D}$, the length scale of the inhomogeneity of the DMI driven spin spiral or domain wall in preferring out-of-plane magnetic moments over in-plane moments, $\ell_\mathrm{ih}=\frac{1}{4\pi}|D/K|$, and   the domain-wall width, $\ell_\mathrm{w}=\frac{1}{\pi}\sqrt{A_{\|}/\abs{K}}$. In this sense $\kappa=\frac{2}{\pi^3}(\ell_\mathrm{hs}/\ell_\mathrm{ih})=\frac{4}{\pi^4}(\ell_\mathrm{hs}/\ell_\mathrm{w})^2$ is a dimensionless parameter relating different length scales. 

\subsubsection{Spin-spiral solution in one dimension}
\label{ssec:1D}
The reduced parameter $\kappa$ is used to determine the micromagnetic texture of the MMLs on the basis of an one-dimensional micromagnetic energy functional [the one-dimensional version  of \eqref{eq:1}].   The lowest-energy ground state in each Co layer is the periodic magnetic N\'eel-type spin spiral $\vc{m}(x)=\sin\Theta(x)\,\vcn{\vc{e}}_x+\cos\Theta(x)\,\vcn{\vc{e}}_z$ for $\kappa \in [0,1)$, with increasing inhomogeneity of the spiral [the change of rotation angle $\Theta(x)$ of the magnetization spiral along the propagation direction $x$ of the spiral]  as $\kappa$ approaches 1, and the  ferromagnetic state for $\kappa > 1 $. A magnetic phase transition under an external magnetic field can be expected when the material has a $\kappa$ close to 1. 

The $\kappa$ values are summarized in  Table~\ref{tab:table} and a small subset is shown in the left panel of  Fig.~\ref{fig:kappa-skyProf}(a). Based on our results, the Ag/Co/Pt and Cd/Co/Pt MMLs show $\kappa$ values of $0.26$ and $0.93$, respectively, thus $\kappa$ values smaller than 1, and their ground state is a spin-spiral state. Since  for the Ag system the $\kappa$ value is small we expect an almost homogeneous [\ie $\Theta(x) \simeq \frac{2\pi}{\lambda}x$] counterclockwise rotating N\'eel-type  spin spiral with a period  $\lambda_\mathrm{ss}\approx\lambda_\mathrm{hs}=-\frac{2A_{\|}}{D}=-\ell_\mathrm{hs}$ of about $-34.6\,$nm and for the Cd system an inhomogeneous analog with $\lambda_\mathrm{ss}=\epsilon \lambda_\mathrm{hs}=-25\,$nm. The minus sign stands for the counterclockwise rotation of the spiral. For $\kappa=0.93$, which is close to $\kappa=1$, the elongation of the pitch of the spiral due to the inhomogeneity becomes an additional factor which is taken care of by the parameter $\epsilon$ which amounts  to about 1.4,  the determination of which involves elliptic integrals (for details see Refs.~\cite{Dzyaloshinskii:65,Zimmermann:14}). The $\kappa$-dependence of the pitch, $\lambda_\textrm{ss}(\kappa)$,  of the spin-spiral texture in units of the domain-wall width was calculated on the basis of the elliptic integrals and is shown as full line in the left panel of Fig.~\ref{fig:kappa-skyProf}(a) connecting the Ag with the Cd system.

For the other MMLs  $\kappa > 1$ holds, thus each trilayer exhibits a collinear ferromagnetic ground state. On top of the ferromagnetic state, meta-stable 180$^\circ$ domain-walls [solution of \eqref{eq:1} with boundary condition $m_z(x=\pm\infty)=\pm 1$ for $K>0$, or $m_x(x=\pm\infty)=\pm 1$ for $K<0$] with a domain-wall width  of  $\ell_\mathrm{w}$ can form as one-dimensional metastable magnetization textures. The domain-wall widths are collected in Table~\ref{tab:table}. We find for all systems domain-wall widths on the nanometer scale. This  is consistent to experimentally measured and theoretically analyzed domain-walls in ultra-thin magnetic films~\cite{Kubetzka:02,Heide:08}. In multilayers  composed of ultrathin magnetic films or ultrathin magnetic films deposited on substrates, interface determined properties dominate, like the magnetocrystalline anisotropy $K_\mathrm{soc}$ dominates over bulk properties such as the spin-stiffness $A$, \ie in thin films, $A$ is relatively small, but $K$ is large which finally leads to small domain-walls. Since the skyrmion size relates to the domain-wall width we expect also small atomic-scale and nanometer-scale skyrmions.

\subsubsection{Skyrmion solution in two dimensions}
\label{ssec:2D}

Isolated chiral skyrmions occur in the ferromagnetic regime as local energy minimizers in a nontrivial homotopy class~\cite{Bogdanov:89,Melcher:14,Li-X:18}. It should be emphasized, since the $Z$/Co/Pt MMLs with $Z=$ Cu, Zn, Y, Zr, Ru, Rh, Pd, Ag, Cd, and Au, \ie 10 of the 13 studied systems, have an out-of-plane easy axis  and a finite DMI, the formation of single chiral skyrmions should be always possible for a well-chosen magnetic field. The films with an in-plane easy axis are chiral in-plane or chiral $XY$-magnets and might be interesting materials for the largely unexplored field of in-plane skyrmions or vortex-antivortex pairs~\cite{Zhang:15.1,moon:19,Zarzuela:20,Kuchkin:20}, subjects beyond the scope of this paper. For Ag/Co/Pt and Cd/Co/Pt no zero-field skyrmion solutions can be  found. A finite field is necessary to straighten the spiral ground state to enable skyrmions. The $Z=$ Cu, Zn, Pd, Au MML systems have  $\kappa$ values close to 1. This makes them  promising MML candidates  hosting skyrmions as metastable states as they can be stable under zero or small external magnetic  fields.

The size of the skyrmion is an important property that  characterizes the skyrmion, determines the choice of experimental means for observation and its role in applications. The skyrmion profile is typically divided into three regions, the core region and the outer region, which are separated by a thin intermediate region, sometimes called the domain-wall region. For the skyrmions we are discussing, in the core region, the skyrmion profile $\Theta(\rho)$ [its exact relation to the magnetization density $\vc{m}(\vc{r})$ is given below] as function of radius $\rho$ from center of the skyrmion at $\rho=0$ to infinity, where it has approached the ferromagnetic background state in form of an exponential tail, has an  arrow-like shape with the steepest slope at the center of the skyrmion. It is customary to work with two definitions of radii: (i) The core radius $R_\mathrm{c}$, $R_\mathrm{c} = R_0 -  \Theta_0(\diff{\Theta}/\diff{\rho})_{|\rho= R_0}^{-1}$ where $(\Theta_0,R_0)$ is the point of inflection of the skyrmion profile~\cite{Hubert:98} or the center of the skyrmion if no inflection point exists.  (ii)  The skyrmion radius $R_\textrm{Sk}$, defined  as the radius at which the $z$ component of the magnetization field changes sign [$m_z(R_\textrm{Sk})=0$ or $\Theta(R_\textrm{Sk})=\pi/2$]. The latter radius definition pays attention to the strong contrast between out-of-plane and in-plane magnetizations reflected by many experimental characterization tools.

The inflection point often indicates a point at which the physical behavior of the magnetization density of a localized micromagnetic object changes. In the case of the skyrmion, the steep linear behavior of the arrow-like shape is separated from the slow exponential behavior. The core radius is obviously a measure for the size of the particle, if the rotation of the magnetization in the region before reaching the  point of inflection  could be continued until the full  rotation of $\pi$. In micromagnetism, the core radius  gives a  measure of the core or the characteristic size of a localized magnetization profile. The skyrmion radius describes the radius at which the magnetization has rotated by 90$^\circ$. If the skyrmion would consist only of the core region with a fast linear change of the profile as a function of the radius over across 180$^\circ$, the skyrmion radius would be just half of the core radius. The deviation from $R_\textrm{c}=2 R_\mathrm{Sk}$ is a measure for the localization of the skyrmion. 

In recent years, considerable progress has been made in the analytical investigation of the stability and size of the skyrmion~\cite{Komineas_small:20,Komineas_large:19,Wang:18,Mantel:18,Buttner:18,Rohart:13,Iwasaki:13,Zhang:15.2,Vidal-Silva:17} as function of micromagnetic parameters. Since the magnetic Co in the trilayer system is of monolayer thickness, we expect atomic-scale skyrmions~ \cite{Romming:15} with radii of a few nanometers or smaller.  In this context, noteworthy is the work of Komineas \etal~\cite{Komineas_small:20} who investigated the profile of skyrmions in the limit of small skyrmions and derived an analytical expression for the skyrmion radius,  $R_\textrm{Sk}$, in the absence of an external magnetic field:
\begin{equation}
R_\mathrm{Sk}[\mathrm{nm}]\approx-\frac{|D|/K}{4\pi\ln\left(\frac{|D|}{\sqrt{A_{\|}K}}\right)}\quad\text{for}\quad  \kappa \gtrsim 40\, .
\label{eq:Rsk_Komineas_small}
\end{equation}
If we apply this expression for the Y($\kappa=24.5$), Zr(2607), Ru(275) and Rh(66.4) system with the $\kappa$ parameters taken from Table~\ref{tab:table}, we obtain tiny skyrmion radii of $R_\mathrm{Sk}(\mathrm{Y})=0.15\,$nm, $R_\mathrm{Sk}(\mathrm{Zr})=0.01\,$nm, $R_\mathrm{Sk}(\mathrm{Ru})=0.02\,$nm and $R_\mathrm{Sk}(\mathrm{Rh})=0.09\,$nm. 

In this paper we follow an alternative route and numerically solve for all $Z$/Co/Pt multilayers with out-of-plane easy axis the skyrmion profile equation~\cite{Bogdanov:94}
\begin{equation}
    {\Theta}^{\prime\prime}  +\frac{1}{\xi}{\Theta}^{\prime} - \frac{1}{2\xi^2} \sin(2\Theta)  + \frac{2}{\xi} \sin^2\Theta - \frac{k}{2} \sin(2\Theta)  - b \sin\Theta =0\, ,
  \label{eq:ODE_reduced}
\end{equation}
with the boundary conditions of a single skyrmion immersed in a ferromagnetic background,
\begin{equation}
\Theta(0)=\pi \quad\text{and}\quad\lim_{\xi \rightarrow \infty}\Theta(\xi)=0\, ,
 \label{eq:BC_ODE_reduced}
\end{equation}
under zero and at finite magnetic  fields $B$ applied in   $z$-direction (normal to the MML plane)  using the shooting method~\cite{Leonov:16,Kiselev:11,Bogdanov:94,Zimmermann:preparation}, whose solution is the skyrmion profile $\Theta(\rho)$ of a single axially symmetric N\'eel-type skyrmion that minimizes the energy of the micromagnetic energy functional \eqref{eq:1}, with the magnetization field $\vc{m}(\vc{r})=\sin{\Theta{(\rho)}}\vcn{\vc{e}}_\rho+\cos{\Theta{(\rho)}}\vcn{\vc{e}}_z$, when expressed in a cylindrical coordinate system $(\rho,\varphi,z)$,    
with the typical relation between  the Cartesian and cylindrical coordinate system, $\vc{r}=(\rho\cos\varphi, \rho\sin\varphi, z)$.
$\vcn{\vc{e}}_\rho$ and $\vcn{\vc{e}}_z$ are the unit vectors in the radial direction and the direction normal to the plane of the MML.
The variables and parameters entering in \eqref{eq:ODE_reduced} are made dimensionless. $\xi = 2\pi \rho / \ell_\mathrm{hs}$ relates to the spatial radius $\rho$ through the length of the homogeneous DMI spiral,  $\ell_\mathrm{hs} = 2A_{\|}/|D|$, and the internal and external fields, $k=K/K_0=(\pi/2)^2\kappa$ and $b=B/(2K_0)$, are scaled relative to the magnetic field $K_0$ required to untwist the one-dimensional spin spiral, $K_0=A_{\|}/\ell_\mathrm{hs}^2 = D^2/(4A_{\|})$. A more detailed description of the method to solution of \eqref{eq:ODE_reduced} with boundary condition \eqref{eq:BC_ODE_reduced} can be found in Ref.~\cite{Zimmermann:preparation}. 

\begin{figure}[t!]
	\centering
	\includegraphics[width=87mm]{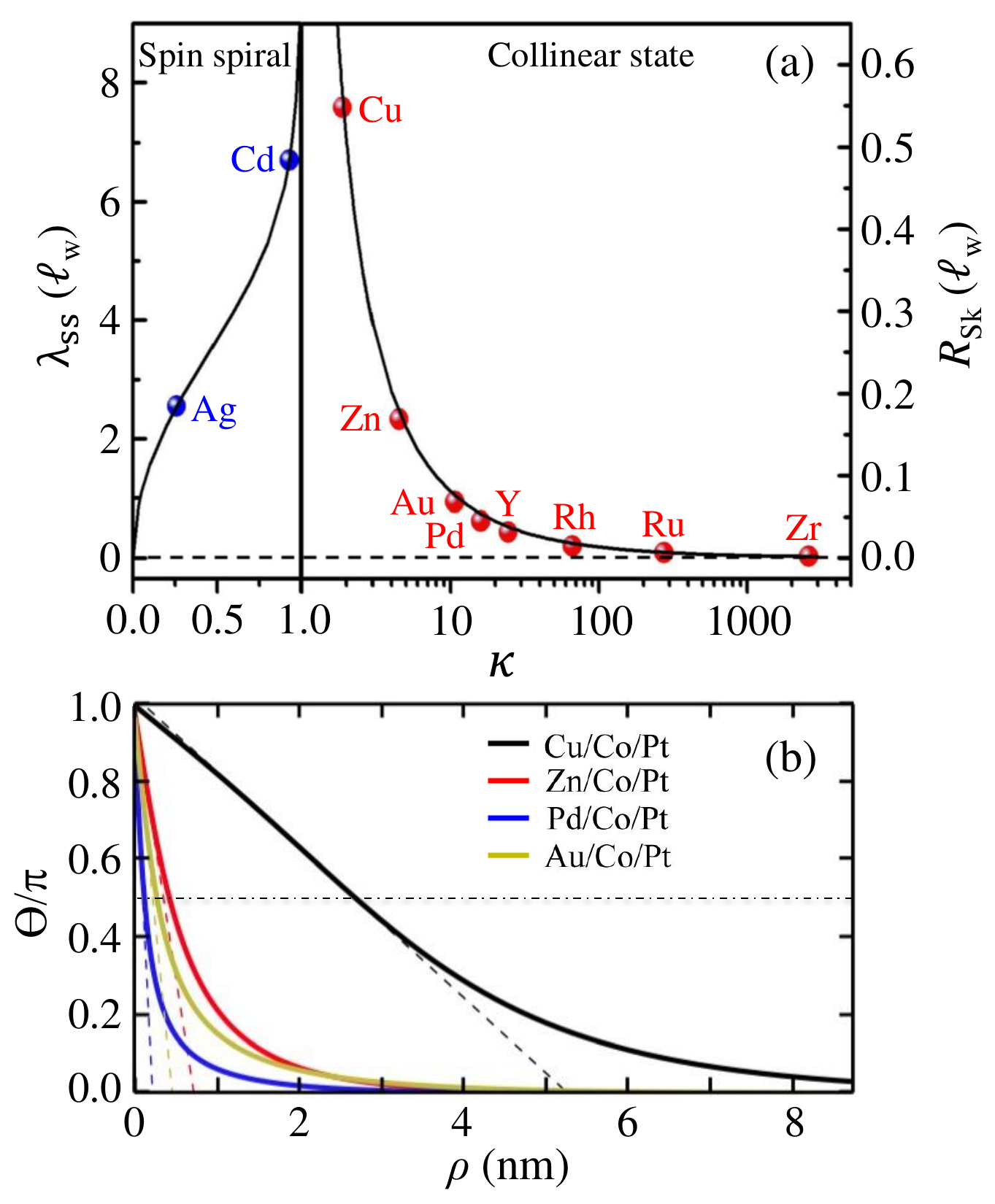}
	\caption{(a) (left panel) Pitch, $\lambda_\mathrm{ss}$, of the spin-spiral ground state and skyrmion radius of the metastable isolated skyrmions (right panel) in units of the domain-wall width, $\ell_\mathrm{w}$, as function of dimensionless micromagnetic parameter $\kappa$ for   $Z$/Co/Pt ($Z=3d$:\@ Cu, Zn; 4$d$:\@ Y, Zr, Ru, Rh, Pd; 5$d$:\@ Au) MMLs. The regimes of collinear state ($\kappa>1$) and spin-spiral state ($\kappa <1$) are shown in the two panels, respectively, ($\kappa\rightarrow$1 indicates a second-order phase transition). The full line in the right panel connecting  Cu to Zr system is the analytical expression of the skyrmion radius \eqref{eq:Rsk_ln_alpha_kappa}. Note the different scales of the left and right panels. (b) Magnetization profile $\Theta(\rho)$ calculated by solving \eqref{eq:ODE_reduced} for an isolating skyrmion using a shooting method.  The intersection point of the dotted line and the $\rho$ axis is the skyrmion core radius $R_\mathrm{c}$. The intersection of the skyrmion profile with the horizontal dashed-dotted line at $\Theta/\pi=0.5$ gives the skyrmion radius $R_\mathrm{Sk}$. }
	\label{fig:kappa-skyProf}
\end{figure}
The skyrmion radius obtained by the  equilibrium solutions, \ie at zero applied $B$ field, of the skyrmion profiles $\Theta(\rho)$ are collected in Table~\ref{tab:table} in units of nanometer  and in the right panel of Fig.~\ref{fig:kappa-skyProf}(a) as function of $\kappa$ in units of the domain-wall width $\ell_\mathrm{w}$. We find that our systems cover a range of three orders of magnitude in $\kappa$ and two orders of magnitude in the domain-wall width normalized skyrmion radius. Inspired by the derivations of Komineas \etal~\cite{Komineas_small:20}, we found  an analytic relation \cite{Zimmermann:preparation} between the skyrmion radius given in units of the domain-wall width  and the dimensionless micromagnetic parameter $\kappa$,
\begin{eqnarray}
    R_\mathrm{Sk}(\kappa)[\ell_\mathrm{w}]&\simeq& \frac{1}{\pi\sqrt{\kappa}}\frac{1}{\ln(\alpha\pi\sqrt{\kappa})}\quad\text{with}\quad\alpha=0.35
    \label{eq:Rsk_ln_alpha_kappa}\\
    &&\qquad \text{for}\quad R_\mathrm{Sk} \lesssim 2 \quad\text{or}\quad \kappa > 1.1 \nonumber, 
\end{eqnarray}
that models all our results very well [see full line in right panel of Fig.~\ref{fig:kappa-skyProf}(a)] for all radii between zero and $0.55\,\ell_\mathrm{w}$, and a $\kappa$ range over three orders of magnitude from  $\kappa=1.88$ to 2607. We see that the skyrmion radius falls off faster than $1/\sqrt{\kappa}$ for large $\kappa$. For  the Y($R_\mathrm{Sk}=0.10\,$nm)  Zr($R_\mathrm{Sk}=0.01\,$nm), Ru($R_\mathrm{Sk}=0.02\,$nm) and Rh($R_\mathrm{Sk}=0.06$~nm) systems we confirm the tiny skyrmions estimated by \eqref{eq:Rsk_Komineas_small}, and for the systems with $\kappa \gtrsim 40$, the agreement is really excellent and can be further improved by introducing a small logarithmic dependence of $\kappa$ on $\kappa$. A similarly small skyrmion radius we obtain for the Zn/Co/Pt($R_\mathrm{Sk}=0.44\,$nm), Pd($0.13\,$nm) and Au($0.29\,$nm) MMLs. 

For $Z$/Co/Pt ($Z=$ Cu, Zn, Pd, Au) the skyrmion profiles $\Theta(\rho)$ of the isolated skyrmions are shown in Fig.~\ref{fig:kappa-skyProf}(b). For  $Z$/Co/Pt ($Z=$ Zn, Pd, Au) MMLs we exhibit skyrmions with very inhomogeneous magnetization profiles expressed in terms of small core radii $R_\mathrm{c}$ of $0.74\,$nm, $0.23\,$nm, and $0.47\,$nm, which is about 1.6 times  the size of  the corresponding skyrmion radii. In the limit of large $\kappa$ and small skyrmion radii, $R_\mathrm{Sk}$, in units of the domain-wall width, the skyrmion profile in the core region is controlled by the exchange energy as the DMI and magnetic anisotropy energy are comparatively small. In this case the skyrmion profile approaches the profile of the  Belavin-Polyakov skyrmion~\cite{Belavin:1975} which is known analytically as $\Theta(\rho)= 2\arctan(R_\mathrm{Sk}/\rho)$. This is an excellent approximation to the profiles in the core region of  the Zn, Pd, and Au systems.  For  Belavin-Polyakov skyrmion the turning point of the magnetization is consistent with our numerical results at the center of the skyrmion at $\rho=0\,$nm and the core radius is $R_\mathrm{c}=(\pi/2)R_\mathrm{Sk}$, close to the value of 1.6 found numerically.  Very small core radii indicate unstable isolated skyrmions with a fast rotation of the magnetization.  The Cu/Co/Pt MML shows in comparison to the other systems a much more homogeneous rotation of the magnetization from the skyrmion core with a core radius of $5.29\,$nm, slightly less than twice the skyrmion radius of $R_\mathrm{Sk}(\textrm{Cu})=2.72\,$nm, and the turning point of the skyrmion profile appears at a finite radius of about 2$\,$nm. 

We can use Eq.~\eqref{eq:Rsk_ln_alpha_kappa} to estimate the temperature dependence of the skyrmion radius assuming that the exchange interactions, the dipolar anisotropy, and the DMI scale approximately with the square of $m = M(T)/M(T=0)$, while the anisotropy $K_\mathrm{soc}$ is proportional to $m^3$. These approximations can be refined for specific cases~\cite{Evans:20}, but here they might suffice to demonstrate the general trends. Splitting $\kappa$ in parts containing $K_\mathrm{dip}$ and $K_\mathrm{soc}$ we see that the former ($\kappa_\mathrm{dip}$) is temperature independent while the latter ($\kappa_\mathrm{soc}$) scales with $m$. Expressing $\kappa(T) = \kappa_\mathrm{dip} + m(T) \kappa_\mathrm{soc}$ or relative to $\kappa(0)$ at $T=0$,
\begin{equation}
\kappa(T) =\kappa(0)-[1-m(T)]\kappa_\mathrm{soc}
\end{equation}
and the domain-wall width $\ell_\mathrm{w} (T) = \frac{1}{\pi}\sqrt{A_{\|}/| K_\mathrm{dip} + m(T)K_\mathrm{soc}|}$ or relative to the width $\ell_\mathrm{w}(0)$ at $T=0$ 
\begin{equation}
 \ell_\mathrm{w} (T) =\ell_\mathrm{w}(0)\sqrt{\frac{|K_\mathrm{soc}+K_\mathrm{dip}|}{|m(T)K_\mathrm{soc}+K_\mathrm{dip}|}}\, ,
\end{equation}
we find that $\kappa(T)$ decreases and the domain-wall width increases with temperature. Both lead to an increase of the skyrmion radius with temperature.
Inserting both into Eq.~\eqref{eq:Rsk_ln_alpha_kappa} we can estimate the change of the skyrmion radius with temperature. Taking Ru/Co/Pt as example where, according to the MC calculations, the magnetization drops at room temperature to half its value, we can estimate that also $\kappa$ is reduced from $274.87$ to $125.38$ and $\ell_\mathrm{w}$ increases from $3.13$ to $4.63$~nm. Accordingly, $R_\mathrm{Sk}$ increases from $0.02$ to $0.04\,$nm  at a temperature that is 88\% of $T_\mathrm{c}$. If a material has a $\kappa$ value close to unity, due to the diverging behavior of Eq.~\eqref{eq:Rsk_ln_alpha_kappa} at this point [Fig.~\ref{fig:kappa-skyProf}(a)] larger effects can be expected: e.g., in the case of Cu/Co/Pt with $T_\mathrm{c}=545\,$K at room temperature $m=0.76$ is found. In this case, $\kappa$ is reduced from $1.88$ to $1.36$, the domain-wall  width $\ell_\mathrm{w}$ is increased from $4.97$ to $5.84\,$nm,  and $R_\mathrm{Sk}$ increases  by about 110\% from $2.72$ to $5.70\,$nm. 

So far we have primarily focused on the size of the skyrmion. The sign of $D$ goes along with the handedness of the magnetic skyrmion. For the noble metal systems (Cu, Ag, Au), the post-transition metals  (Zn, Cd), the early (Y), and the late transition metals (Pd), as well as Ru, $D$ is larger than zero and we expect skyrmions with a magnetic texture of counterclockwise handedness following the magnetization inside out.  The systems in the first half of the $4d$ transition-metal series (Nb, Mo, Tc) have negative $D$ and negative $K$ and thus they do not form chiral skyrmions.  Only the Zr and Rh exhibit $D<0$ and positive $K>0$ and can form skyrmions with a clockwise magnetization texture. 

Two interesting systems not discussed  so far are the  Ag/Co/Pt and Cd/Co/Pt MMLs, for which no skyrmions can be found at zero field and which have been discussed in Sec.~\ref{ssec:1D} to exhibit the one-dimensional spin-spiral  ground state, a magnetization texture of  counterclockwise rotational sense. For external magnetic fields, $B$,  larger than $B\gtrsim K_0=D^2/(4A_{\|})$, which amounts to $0.29$ and $0.24\,$T for the cases of Ag and Cd MMLs, respectively, a phase transition to a two-dimensional hexagonal skyrmion lattice phase of counterclockwise rotating skyrmions  with a lattice constant of about
$\lambda_\mathrm{is}(\mathrm{Ag})=25.0$~nm and $\lambda_\mathrm{hs}(\mathrm{Cd})=34.6$~nm   can be expected~\cite{Bogdanov:99}. The lattice constant has a weak dependence on the external field, which is negligible on the level of discussion here. Releasing the external field  adiabatically to zero field can freeze in the skyrmion lattice phase as metastable phase at zero field. Single skyrmions of the size of about $\lambda_\mathrm{h(i)s}$ can emerge at fields close to the transition to the field-polarized ferromagnetic state.  

Looking at the right panel at Fig.~\ref{fig:kappa-skyProf}(a) or at expression \eqref{eq:Rsk_ln_alpha_kappa} we observe that the skyrmion radius diverges to infinity  if $\kappa$ approaches 1. The  behavior of the skyrmion radius as function of $\kappa$ is not properly described by \eqref{eq:Rsk_ln_alpha_kappa} and we suggest the following  expression~\cite{Zimmermann:preparation}:
\begin{eqnarray}
    R_\mathrm{Sk}(\kappa)[\ell_\mathrm{w}]&\simeq& \frac{1}{2\sqrt{\kappa}}\frac{1}{\W(\alpha\pi\sqrt{\kappa-1})}\,\,\,\text{with}\,\,\,\alpha=0.35
    \label{eq:Rsk_W_alpha_kappa}\\
    &&\qquad\quad\,\, \text{for}\quad R_\mathrm{Sk} \gtrsim 1 \quad\text{or}\quad \kappa \lesssim 1.6\, , \nonumber 
\end{eqnarray}
to describe the skyrmion radius in the limit of large skyrmions. $\W$ is the upper branch of Lambert W function~\cite{Hoorfar:08}. For $\kappa < 1+ 1/(\alpha\pi \mathrm{e})^2\approx 1.112$, the inverse W function $\mathrm{W}_0^{-1}(x)$, with $x:=\alpha\pi\sqrt{\kappa-1}$, can be approximated by $\mathrm{W}_0^{-1}(x)\approx\frac{1}{x} + 1 -\frac{1}{2}x$. Thus, the skyrmion radius diverges to infinity with the leading asymptotic behavior  proportional to  $R_\mathrm{Sk}\propto  1/\sqrt{\kappa -1}$  if $\kappa$ approaches unity, consistent with the expansion of the skyrmion radius in the limit of  large skyrmions by  Komineas \etal~\cite{Komineas_large:19}.

\subsubsection{Prediction towards systems with 10-nm skyrmions}
\label{ssec:Rsk-10}
Using our analytical model \eqref{eq:Rsk_ln_alpha_kappa} for the skyrmion radius as a function of $\kappa$, we estimate at which Co layer thickness $t_{N\mathrm{Co}}$  or at how many Co layers $N_\mathrm{Co}$ of the monatomic layer thickness $t_{1\mathrm{Co}}=0.2\,$nm, skyrmions with a diameter of $10\,$nm can be expected.  In Table~\ref{tab:table} the micromagnetic parameters have been given per formular units.  Considering that the spin stiffness $A_{\|}$ and the shape anisotropy $K_\mathrm{dip}$ are quantities that depend on the volume of the material, the spin-orbit contribution to the magnetic anisotropy $K_\mathrm{soc}$ and the DMI are in first approximation interface properties, and if we consider the energy density in terms of the energy per volume, then the corresponding micromagnetic parameters $A_{\|}$ and $K_\mathrm{dip}$ remain  independent of the thickness, but $K_\mathrm{soc}$ and $D$ scale  inversely proportional with the Co film thickness, \ie $D\rightarrow D/N_\mathrm{Co}$ and $K_\mathrm{soc}\rightarrow K_\mathrm{soc}/N_\mathrm{Co}$. Consequently we can change $\kappa(N_\mathrm{Co})$ with the number of Co layers according to  
\begin{eqnarray}
\kappa(N_\mathrm{Co})
&=&  \left(\frac{4}{\pi}\right)^2\frac{A_{\|} (K_\mathrm{soc}+N_\mathrm{Co}K_\mathrm{dip})}{D^2}N_\mathrm{Co} \label{eq:kappa_of_Nco} \\
&=& \kappa{(1)}N_\mathrm{Co}+ \kappa_\mathrm{dip}N_\mathrm{Co}(N_\mathrm{Co}-1)\, ,
    \label{eq:kappa_of_t}
\end{eqnarray}
with $\kappa=\kappa(1)=\kappa_\mathrm{soc}+\kappa_\mathrm{dip}$ and $\kappa_\mathrm{dip}=(4/\pi)^2 A_{\|}K_\mathrm{dip}/D^2$. Accordingly, we can choose the  thickness of the Co-layer so that  $\kappa(N_\mathrm{Co})$ can be set close to $\kappa(N_\mathrm{Co})=1$, where the skyrmion radius diverges like $R_\mathrm{Sk}\propto 1/\sqrt{\kappa-1}$ and thus any desired skyrmion radius can be set. We observe that the first term of \eqref{eq:kappa_of_t} increases $\kappa(N_\mathrm{Co})$ linearly with the number of  Co layers, but the second term reduces $\kappa(N_\mathrm{Co})$ quadratically, although the prefactor $\kappa_\mathrm{dip}$ is much smaller than $\kappa$. A different way of interpretation follows from \eqref{eq:kappa_of_Nco}. Since $K_\mathrm{soc}>0$ and $K_\mathrm{dip}<0$, with increasing Co thickness we approach at $K_\mathrm{soc}+ N_\mathrm{Co}K_\mathrm{dip}=0$ the  reorientation transition from out-of-plane to in-plane easy axis and in this vicinity, $\kappa(N_\mathrm{Co})$ approaches   $\kappa(N_\mathrm{Co})\approx 1$. Obviously, Co layers below a coverage of a monatomic layer, \ie $0< N_\mathrm{Co}<1$, also lead to $\kappa$ values close to one, but this is an issue related to Co dusts at interfaces and is  beyond the scope of this paper, which focuses  on Co-based multilayers. 

Inverting \eqref{eq:kappa_of_t} we determine the number of Co layers for a chosen value $\kappa(N_\mathrm{Co})$  by
\begin{eqnarray}
  N_\mathrm{Co} &=& \frac{1}{2} \left(1-\frac{\kappa(1)}{\kappa_\mathrm{dip}}\right) \pm   \frac{1}{2} \sqrt{\left(1-\frac{\kappa(1)}{\kappa_\mathrm{dip}}\right)^2\!+4\frac{\kappa(N_\mathrm{Co})}{\kappa_\mathrm{dip}}} 
  \label{eq:NCo_kappaN}\\
  &\simeq& \frac{K_\mathrm{soc}}{|K_\mathrm{dip}|} -\frac{\kappa(N_\mathrm{Co})}{\kappa_\mathrm{soc}} \quad\text{for}\quad K_\mathrm{soc}\gg |K_\mathrm{dip}|,\,  N_\mathrm{Co}\ge 1\nonumber \, .
\end{eqnarray}
Let us denote $\bar{N}_\mathrm{Co}$ as the critical layer thickness by which the skyrmion radius diverges, \ie defined by the fact that $\kappa(\bar{N}_\mathrm{Co})=1$. Since according to Table~\ref{tab:anisotropy-K}, $K_\mathrm{soc}/|K_\mathrm{dip}|\approx 8$ to 32, we expect layer thicknesses in the order of  $N_\mathrm{Co} \approx 7$ to 31 to come close to $\kappa(\bar{N}_\mathrm{Co})=1$. For example, for the Cu system $K_\mathrm{soc}/|K_\mathrm{dip}|\approx 8$ and $\kappa_\mathrm{soc}=2.17$, thus one expects the divergence of the skyrmion radius between $7$ and $8$ layers. For the Y system we expect thicknesses around 20 to 21 layers. 

At the same time, the domain-wall width increases according to 
\begin{equation}
    \ell_\mathrm{w}(N_\mathrm{Co})=  \frac{1}{\pi}\sqrt{\frac{A_{\|}}{K_\mathrm{soc}/{N_\mathrm{Co}}+K_\mathrm{dip}}}\, ,
    \label{eq:lw_of_t}
\end{equation}
for $K_\mathrm{soc}> {N_\mathrm{Co}}|K_\mathrm{dip}|$, diverging at the thickness of the reorientation transition balancing shape anisotropy against spin-orbit induced interface anisotropy. With the number of Co layers chosen according to \eqref{eq:NCo_kappaN}, the domain-wall width increases to 
\begin{equation}
    \ell_\mathrm{w}(N_\mathrm{Co})\simeq  \ell_\mathrm{w}(1) \frac{K_\mathrm{soc}}{|K_\mathrm{dip}|}\sqrt{\frac{\kappa(1)}{\kappa(N_\mathrm{Co})}}  \quad\text{for}\quad K_\mathrm{soc}\gg |K_\mathrm{dip}|\, .
    \label{eq:lw_of_NCo}
\end{equation}
Considering the ratios $K_\mathrm{soc}/|K_\mathrm{dip}|$ of 8 to 32 we have obtained for our systems, we easily gain an additional order of magnitude in skyrmion size.  

Using the relations  \eqref{eq:kappa_of_t} and  \eqref{eq:lw_of_t} in the analytical models \eqref{eq:Rsk_ln_alpha_kappa} and \eqref{eq:Rsk_W_alpha_kappa} we conclude on the following expressions for  the skyrmion radii in the limit of  small radii [$R_\mathrm{Sk}(N_\mathrm{Co})/\ell_\mathrm{w}(N_\mathrm{Co})<2$]   
\begin{equation}
    R_\mathrm{Sk}(N_\mathrm{Co})\simeq \frac{1}{\pi \sqrt{\kappa(N_\mathrm{Co})} }\frac{1}{\ln[\alpha\pi\sqrt{\kappa(N_\mathrm{Co})}]}\, \ell_\mathrm{w}(N_\mathrm{Co})\, ,
    \label{eq:Rsk_ln_alpha_kappaN}
\end{equation}
and of large radii [$R_\mathrm{Sk}(N_\mathrm{Co})/\ell_\mathrm{w}(N_\mathrm{Co})>1$] 
\begin{equation}
    R_\mathrm{Sk}(N_\mathrm{Co})\simeq \frac{1}{2\sqrt{\kappa(N_\mathrm{Co})}}\frac{1}{\W[\alpha\pi\sqrt{\kappa(N_\mathrm{Co})-1}]}\, \ell_\mathrm{w}(N_\mathrm{Co})
    \label{eq:Rsk_W_alpha_kappaN}\\
\end{equation}
in units of nanometer with respect to the number of Co-layers. According to \eqref{eq:kappa_of_t},  $\kappa$  behaves in the vicinity $\Delta {N}_\mathrm{Co}$ of the critical thickness $\bar{N}_\mathrm{Co}$, $\Delta {N}_\mathrm{Co}=\bar{N}_\mathrm{Co}-{N}_\mathrm{Co}$, as $\kappa({N}_\mathrm{Co})=\kappa(\bar{N}_\mathrm{Co}-\Delta{N}_\mathrm{Co})\simeq 1 + \kappa_\mathrm{soc}\Delta{N}_\mathrm{Co}$, and considering the expansion of the W function around $\Delta {N}_\mathrm{Co}=0$ we expect that the skyrmion radius 
\begin{eqnarray}
R_\mathrm{Sk}(\Delta {N}_\mathrm{Co})  &\approx & \frac{1}{2}\frac{\sqrt{\kappa_1}}{1 + \kappa_\mathrm{soc}\Delta{N}_\mathrm{Co}}\left(\frac{1}{\pi\alpha\sqrt{\kappa_\mathrm{soc}}\sqrt{\Delta{N}_\mathrm{Co}}}+1\right) \nonumber \\ 
&& \times \frac{K_\mathrm{soc}}{|K_\mathrm{dip}|}\ell_w(1) \\
&& \text{for}\quad 0\le \Delta{N}_\mathrm{Co} < \frac{1}{\pi^2\alpha^2e^2\kappa_\mathrm{soc}} \nonumber
\end{eqnarray}
diverges in the vicinity $\Delta {N}_\mathrm{Co}$ according to $R_\mathrm{Sk} \propto 1/\sqrt{\Delta {N}_\mathrm{Co}}$ or $R_\mathrm{Sk} \simeq R_0 + R_1/\sqrt{\bar{N}_\mathrm{Co}-{N}_\mathrm{Co}}$ for ${N}_\mathrm{Co}< \bar{N}_\mathrm{Co}$, respectively, where $R_0$ and $R_1$, are parameters taking care on the different prefactors.

\begin{figure}[t]
	\centering
    \includegraphics[width=85mm]{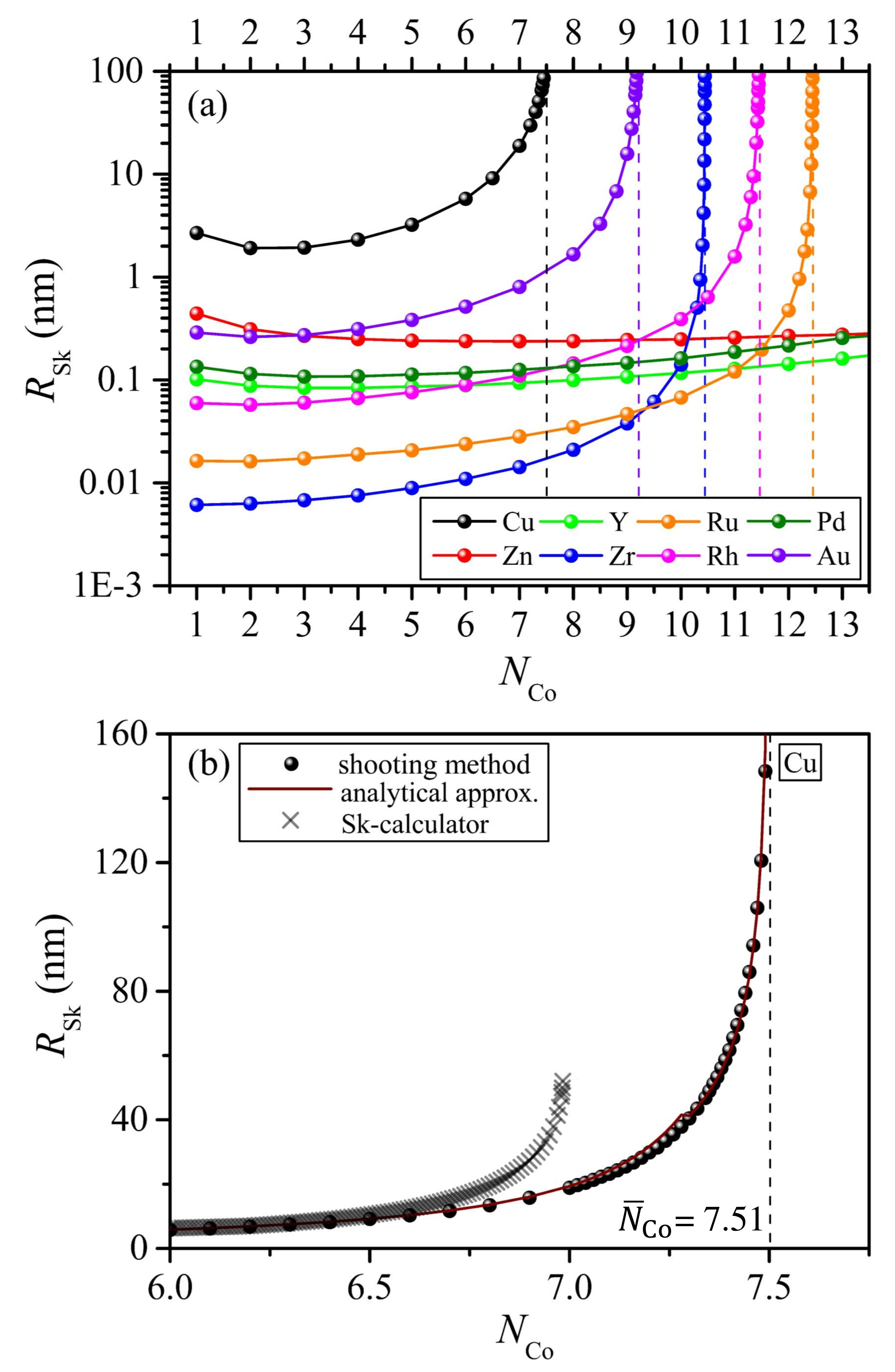}
	\caption{(a) Skyrmion radius as function of the number of Co layers for $Z$/Co/Pt ($Z=$Cu, Zn, Y, Zr, Ru, Rh, Pd, Au) MMLs. Each little sphere represents a solution of \eqref{eq:ODE_reduced}. Lines are guides to the eyes. (b) Enlarged section of (a) focusing on the Cu system.  Notice the different scale. The spheres represent the results of (a), the full lines represent to analytical approximations \eqref{eq:Rsk_ln_alpha_kappaN} and \eqref{eq:Rsk_W_alpha_kappaN}, and the crosses represent the results obtained using the skyrmion radius calculator~\cite{Zimmermann:preparation,Skyrmion-R-calculator}, taking into account the magnetic film charge to the magnetostatic self-energy, or demagnetization energy, respectively.} 
	\label{fig:Rsk-NCo}
\end{figure}

The results of the skyrmion radius as function of the Co layers obtained by solution of skyrmion profile equation  \eqref{eq:ODE_reduced} are summarized in Fig.~\ref{fig:Rsk-NCo}(a) for the skyrmion forming systems. We limit ourselves to thicknesses of maximum 13 Co layers and radii smaller than $100\,$nm. Fig.~\ref{fig:Rsk-NCo}(a) confirms once again that by selecting a third monatomic layer component $Z$, the skyrmion radius can be varied by a factor of 250 already for  systems with only one Co layer.  In general, we see that the skyrmion radius increases with the number of Co layers. However, this increase is not detectable for the Y, Pd, and Zn systems within the first 13 layers.  For these systems the ratio $K_\mathrm{soc}/|K_\mathrm{dip}|$ is very large [$Z(K_\mathrm{soc}/|K_\mathrm{dip})$: Y(21), Pd(19), Zn(32)]. If we restrict Co to 13 layers, we obtain the largest increase in radius as function of Co thickness for the Cu, Au, Zr, Rh, and Ru systems with the critical Co thickness of $\bar{N}_\mathrm{Co}=7.51, 9.22, 10.44, 11.47, 12.45$ layers, respectively,  of a divergent skyrmion radius indicated by straight broken lines in Fig.~\ref{fig:Rsk-NCo}(a).
We find that close to the critical Co layer thickness varying the Co thickness by one layer can alter the skyrmion radius easily by  orders of magnitude. This means that in this growth range, the skyrmion radius can be greatly influenced by minute changes in the variation of the film thickness, or minute variations in the growth-induced properties, such as the variation of the strain, simply properties that modify the magnetic anisotropy on a very fine scale. 

The Cu and Au systems host counterclockwise skyrmions that couple ferromagnetically across the Cu and Au interlayers. Skyrmions of $10$-nm diameter are expected for Co thicknesses of $N_\mathrm{Co}(\textrm{Cu})=5.8130$ and $N_\mathrm{Co}(\textrm{Au})=8.6938$ layers. We are fully aware that a 4 decimal accuracy cannot be realized experimentally, but expresses the accuracy required by the shooting method to determine the thickness if the $10$-nm diameter criteria is met in the thickness regime where the skyrmion radius rises strongly. In Fig.~\ref{fig:Rsk-NCo}(a) we note also three promising systems, the Zr, Ru, and Rh system, that can host skyrmions of $10$-nm diameter  coupling antiferromagnetically across the spacer layer, however, for larger Co thicknesses estimated to $N_\mathrm{Co}(\textrm{Zr})= 10.4232$ layers for Zr, $N_\mathrm{Co}(\textrm{Ru})= 12.3860$, and $N_\mathrm{Co}(\textrm{Rh})= 11.2763$ layers for the Ru and Rh system, respectively.  In the Zr and Rh systems, the  magnetic textures of the skyrmions exhibit a clockwise  and in the Ru system a counterclockwise handedness.

Figure~\ref{fig:Rsk-NCo}(b) provides a detailed view of the Cu  system.  We  observe how the radius shows the expected $1/\sqrt{\Delta {N}_\mathrm{Co}}$ behavior for thicknesses larger than 6.5 layers and diverges for layer thicknesses between  7 and 8. Taking the micromagnetic parameters from Tables~\ref{tab:table} and \ref{tab:anisotropy-K}, we determine for the Cu  system the ratio $K_\mathrm{soc}/|K_\mathrm{dip}|=7.996$, $\kappa_\mathrm{dip}=-0.27131$, $\kappa_\mathrm{soc}=\mathrm{2.16953}$  and obtain according to \eqref{eq:NCo_kappaN} the critical films thickness $\bar{N}_\mathrm{Co}=7.5053$  which is in excellent agreement with results obtained by the shooting method. We show that the analytical skyrmion radius model \eqref{eq:Rsk_ln_alpha_kappaN} represents well the results of the shooting method  for a Co thickness up to 7.25  layers for the Cu  system and the model \eqref{eq:Rsk_W_alpha_kappaN} for thickness beyond. At the thickness of the transition from \eqref{eq:Rsk_ln_alpha_kappaN} to \eqref{eq:Rsk_W_alpha_kappaN} one observes a tiny discontinuity in the skyrmion radius giving an account of the error of the approximations.  

Figure~\ref{fig:Rsk-NCo}(b) contains also results (indicated by $\scriptstyle{\mathbf{\times}}$) of the skyrmion radius as function of the number of Co layers with an improved treatment of the magnetostatic self-energy, or demagnetization energy, respectively, beyond the shape approximation to the magnetocrystalline anisotropy. The treatment is improved by the missing film-charge contribution of the magnetization, $\sigma(\vc{r}) = \nabla \cdot \vc{m}_\parallel$, which together with the shape approximation gives the exact contribution of the magnetostatic energy under the condition that the magnetization profile of the skyrmion does not vary along the direction of the film thickness and the magnetization is well described by the continuum model. The calculation of the magnetostatic energy due to the film-charge contribution and the associated determination of the skyrmion radius  goes beyond the scope of this paper and we have carried out the calculation using our recently developed skyrmion radius calculator~\cite{Zimmermann:preparation,Skyrmion-R-calculator}. Since the film-charge contribution to the demagnetization energy prefers to reduce the divergence of the in-plane magnetization, this contribution favors larger skyrmions with a larger portion of in-plane magnetization. Thus, due to this correction, the critical Co thickness reduces from $\bar{N}_\mathrm{Co}=7.51$ layers by about 0.6  layers to $\bar{N}_\mathrm{Co}=6.89$, \ie the skyrmion radius diverges already for a thinner Co-layer. For skyrmion radius larger than $52\,$nm, the skyrmion was not stable anymore.  As a general remark, this can make the exact prediction of the Co thicknesses for $10$-nm skyrmions difficult if the $5$-nm radius lies already in the steep regime  as other approximations to the magnetostatic self-energy, \eg the exact dipole sum versus continuum model, will further change the critical thickness.

The model provides an  estimate  of the skyrmion sizes as function of the layer thickness. Note, the model is based on several assumptions: (i) We assume that the change in Co thickness does not change the electronic structure or the electronic properties to such an extent that the micromagnetic parameters change with Co thickness. 
(ii) For an infinite 2D film, the magnetostatic self-energy (averaged by the film thickness) decomposes to leading orders into a shape anisotropy energy
and a magnetic film charge term, which we neglected and which in reality increases with the number of Co layers, and (iii) we assume a skyrmion tube behavior across the Co film, which in return limits the thickness in the number of Co layers to fulfill this condition. This is the reason why we focused on Co films of maximal 13 layers thickness.  The validity of this model will be further  investigated in the future.

\section{Summary and Conclusion}
We investigated the magnetic properties of (111)-oriented monatomically thick $Z$/Co/Pt ($Z=4d$ series:  Y--Pd, the noble metals: Cu, Ag, Au, the post-noble metals: Zn and Cd)   multilayers by means of density functional theory calculations. We explored the role of the non-magnetic elements $Z$ in modifying the magnetic exchange interactions, especially the interlayer exchange coupling between the magnetic Co layers, the magnetic anisotropy, and the Dzyaloshinskii-Moriya interaction  at the Co/Pt interface. 

The results witness clear chemical trends of the different properties. The apparent size of the $4d$ atoms across the transition-metal series follows the band-filling rule with the smallest atom at the center of the series (Tc, Ru). The size of $4d$ atoms not only determines the $4d$-Pt and $4d$-Co interlayer distances, but also slightly changes the Co-Pt interlayer distance. Across the noble metals and the late  $4d$ transition-metal Pd,  we find a ferromagnetic interlayer coupling, for the post-noble metals as well as for most $4d$ transition metals, with the exception of Mo, we find an antiferromagnetic interlayer coupling.  The exchange interaction within the Co layer is ferromagnetic for all systems, but the spin-stiffness and the magnetic moment of Co changes across the transition-metal series alike from smaller values at the middle  of the early $4d$ series, Nb/Co/Pt, to end  of the series, Pd/Co/Pt. The Rh/Co/Pt system shows the largest spin stiffness and highest induced magnetic $Z$ moment due to the strong hybridization between Co and Rh. The critical temperature follows the trend of the spin stiffness. Across the $4d$ transition-metal series the smallest values are obtained in the middle of the first-half series but increase then again to the end of the series, \ie Cd. With the exception of Nb, Mo, and Tc systems the critical temperatures are above room temperature. The largest critical temperature we estimated for the Rh system is $810\,$K.  All systems have an easy axis of the magnetization that is out of plane with the exception of the early $4d$ transition-metal systems with $Z=$ Nb, Mo, and Tc. They are not candidates for chiral axial skyrmions but may be candidates for in-plane skyrmions. The $Z$ elements modify the sign and magnitude of Dzyaloshinskii-Moriya interaction that varies in the range of $-2.8$ to $11.44\,$meVnm/f.u.. For the noble and post-noble metals, the early ($Z=\,$Y) and late ($Z=\,$Pd) transition as well as Ru, the sign of the  Dzyaloshinskii-Moriya interaction favors a spin texture with a counterclockwise handedness. The systems with $Z$ elements from the first half of the transition-metal series ($Z=$ Zr, Nb, Mo, Tc) as well Rh favor spin textures with clockwise handedness.    

Taking advantage of the concepts of multiscale modeling, we have taken the micromagnetic parameters obtained by DFT and used these parameters in a micromagnetic energy functional. We investigated  the spin texture in the planes of the multilayers on a larger scale by minimizing the materials-specific micromagnetic energy functional in one and two dimensions to explore the formation and stability of N\'eel-type spin-spiral ground states and metastable skyrmions. We numerically solved the skyrmion profile equation using a shooting method to study the skyrmion profile and size. In summary, with the exception of the Cu/Co/Pt multilayers, for $Z$/Co/Pt multilayers with $Z=$ Y, Zr, Ru, Rh, Zn, Pd, Au we find tiny skyrmions with sizes comparable with the atomic lattice. At this limit, both the concept of skyrmions, which assumes a smooth magnetization profile, and the underlying micromagnetic theory break down.  Such tiny skyrmions feel the coarse grain of the underlying crystal lattice and have negligible lifetimes. Therefore, we do not consider them any further.  Despite these limitations, the skyrmion radius estimated by the analytical model of Komineas \etal \cite{Komineas_small:20} agrees very well  with our results obtained for tiny skyrmions  using the shooting method. The Ag/Co/Pt and Cd/Co/Pt multilayers exhibit a one-dimensional spin-spiral  ground state of  counterclockwise rotational sense with a period length of $25$ and $35\,$nm, respectively, that couple ferromagnetically and antiferromagnetically across the Ag and Cd space layer, respectively.  

We found an analytical expression for the skyrmion radius that covers all our numerical results and is valid for a large range of micromagnetic parameters.  Based on this expression (i) we  estimated the skyrmion radius as function of temperature and found that the skyrmion radius may easily double when going from cryogenic to room temperature. (ii) We proposed  a model that allows to extrapolate the skyrmion radius from the \abinitio\ results of multilayers with monatomic films on multilayers with Co films consisting of several atomic layers containing skyrmions, which then can host skyrmions with a diameter of  $10\,$nm.  We expect Cu/Co$_{5.8}$/Pt and Au/Co$_{8.7}$/Pt as interesting  material systems for  ferromagnetically  and  Zr/Co$_{10.4}$/Pt, Ru/Co$_{12.5}$/Pt, Rh/Co$_{11.5}$/Pt  for antiferromagnetically coupled skyrmion systems, \ie in synthetic antiferromagnets. All systems host  counterclockwise skyrmions with the exception of the Zr and Rh systems. We found  that in the vicinity of a critical Co-layer thickness the skyrmion radius changes very rapidly and thus  can be greatly influenced by minute changes in the variation of the film thickness, or minute variations in the growth induced properties that may change the magnetic anisotropy on a very fine scale. 

In the future we want to further validate and improve the model because in combination with high-throughput calculations using density function theory it could eventually open a path to predict optimal multilayers for skyrmions with certain properties, since it allows a fast search in the five-dimensional space spanned by the spin stiffness, magnetic anisotropy, Dzyaloshinskii-Moriya interaction, magnetization and layer thickness. 

\textit{Note added in proof.} Recently, we became aware of a theoretical study by Bernand-Mantel \etal [106, 107] providing analytical expressions of the skyrmion radius of compact magnetic skyrmions beyond DMI by including the higher orders of the dipolar fields.	Since the limit of compact magnetic skyrmions, skyrmions with profiles close to a Belavin-Polyakov profile and dominated by the exchange energy, applies to our Ru, Rh, and Zr/Co/Pt(111) monolayer systems, these analytical expressions may be used to refine our results. For these three systems we find an increase of the skyrmion radius from 0.0162 to 0.0278 nm (Ru), from 0.0596 to 0.1235 nm (Rh) and a decrease from 0.0059 to 0.0042 nm (Zr) due to higher order contributions to the dipolar fields.

\section{Acknowledgments}

We thank Fabian Lux and Christof Melcher for enlightening discussions on the analytic expressions of the skyrmion radius. We gratefully acknowledge financial support from the DARPA TEE program through grant MIPR (Grant No. HR0011831554) from DOI, from the European Union H2020-INFRAEDI-2018-1 program (Grant No. 824143, project "MaX - materials at the exascale"), from the European Research Council (ERC) under the European Union's Horizon 2020 research and innovation program (Grant No.\ 856538, project "3D MAGiC"),  from Deutsche For\-schungs\-gemeinschaft (DFG) through SPP 2137 "Skyrmionics" (Project BL 444/16), the Collaborative Research Centers SFB 1238 (Project C01) as well as computing resources on the supercomputer JURECA at the J\"ulich Supercomputing Centre and on the JARA partition of CLAIX at the RWTH Aachen University under projects jias1f and jara0197, respectively.

\end{document}